\documentclass[%
reprint,
showkeys,
superscriptaddress,
preprintnumbers,
nofootinbib,
amsmath,amssymb,
aps,
prd,
floatfix,
]{revtex4-2}

\usepackage{afterpage}
\usepackage{subfigure}
\usepackage{rotating} 
\usepackage{booktabs}
\usepackage{graphicx}
\usepackage{dcolumn}
\usepackage{bm}
\usepackage{float}
\usepackage{amsmath}
\usepackage{natbib}
\usepackage{hyperref}
\renewcommand{\arraystretch}{1.5}
\usepackage{xcolor}
\newcommand{\reply}[1]{\textcolor{black}{ #1}}

\begin{document}

\preprint{J.-C.~Wang, H.~Song, H.~Li, J.Zhu \& B.-Q.~Ma, \href{https://doi.org/10.1103/748y-3vyj}{Phys. Rev. D 114 (2026) 063023} \href{http://arxiv.org/abs/2608.21266}{[arXiv:2608.21266]}}

\title{Generation of TeV Photons by PeV Neutrinos in Dense Astrophysical Environments}
\author{Jun-Chen Wang}\email{junchenwang@stu.pku.edu.cn}
\affiliation{Department of Physics and Jockey Club Institute for Advanced Study,
The Hong Kong University of Science and Technology, Hong Kong S.A.R., China}
\affiliation{School of Physics, Peking University, Beijing 100871, China}

\author{Hanlin Song}\email{hanlin@stu.pku.edu.cn}
\affiliation{Center for Gravitational Wave Experiment, National Microgravity Laboratory, Institute of Mechanics, Chinese Academy of Sciences, Beijing 100190, China}
\affiliation{School of Physics, Peking University, Beijing 100871, China}

\author{Hao Li}\email{haolee@cqu.edu.cn}
\affiliation{Department of Physics, Chongqing University, Chongqing 401331, China}

\author{Jie Zhu}\email{jiezhu@cqu.edu.cn}
\affiliation{Department of Physics, Chongqing University, Chongqing 401331, China}

\author{Bo-Qiang Ma}\email{mabq@pku.edu.cn}
\affiliation{School of Physics, Zhengzhou University, Zhengzhou 450001, China}
\affiliation{School of Physics, Peking University, Beijing 100871, China}
\begin{abstract}
Recent observations by IceCube and KM3Net of PeV-scale ultra-high-energy (UHE) neutrinos, together with detections of TeV-PeV photons from various sources such as the Crab Nebula, the Galactic Center, and gamma-ray burst  by ground-based observatories including Tibet AS$\gamma$, MAGIC, Carpet-3, and LHAASO, point to the existence of extreme astrophysical environments capable of accelerating particles to ultra-high energies. These findings motivate investigations of possible connections between UHE neutrinos and photons in such environments. Theoretically, dense regions surrounding compact objects can efficiently produce UHE neutrinos. In this work, we calculate the production of UHE photons from neutrino-nucleon interactions, and note that if these interactions occur in the outer, optically thin regions of dense environments, the resulting photons could potentially be observed. In our model, an incident neutrino scatters off a nucleon, generating secondary partons that hadronize into pions and subsequently decay into UHE photons. We calculate the resulting photon energy spectra and find that for incident (anti)neutrinos with energies above 1 PeV, the probability of producing photons with energies exceeding 1 TeV is greater than 13\%.
As a concrete application, we show that this mechanism can quantitatively account for the preburst TeV photons observed in GRB~221009A, providing a natural explanation for both their energies and lead times.
These findings establish a plausible mechanism linking UHE neutrino events to gamma-ray observations, providing new insights into hadronic processes in extreme astrophysical environments and supporting multi-messenger astronomy studies.
\end{abstract}

\maketitle

\section{Introduction}
Ultra-high-energy (UHE) neutrinos and photons serve as critical messengers for probing extreme astrophysical environments that can accelerate particles to energies far beyond terrestrial accelerators. Recent breakthrough detections have expanded our understanding of these cosmic accelerators: the KM3Net Collaboration reported a 220 PeV neutrino,
the highest-energy neutrino ever observed \cite{KM3NeT:2025npi}, following earlier TeV–PeV neutrino measurements by IceCube~\cite{IceCube:2013low, IceCube:2014stg}. Parallel advances in gamma-ray astronomy have yielded equally striking results, including a 300 TeV photon from GRB 221009A (Carpet-3)~\cite{Carpet-3Group:2025fcs}, a 1.4 PeV Galactic photon (LHAASO)~\cite{LHAASO:2021gok}, and TeV–PeV emissions detected by Tibet AS$\gamma$~\cite{TibetASg:2019ivi} and MAGIC~\cite{MAGIC:2022iap}. Together, these observations confirm the existence of cosmic accelerators capable of reaching extreme energy scales, yet the underlying particle acceleration and emission mechanisms remain incompletely resolved \cite{Arguelles:2024ncf}.

In extreme astrophysical settings such as gamma-ray burst (GRB) progenitors~\cite{Murase:2013ffa}, flat-spectrum radio quasars~\cite{Dermer:2014vaa}, and core-collapse supernovae~\cite{Bykov:2015nta, Chakraborty:2015sta, Petropoulou:2017ymv}, UHE particles are generated via distinct processes. UHE neutrinos typically arise from hadronic interactions: accelerated protons or heavy nuclei collide with ambient matter or radiation fields, triggering proton–proton ($pp$) or proton–photon ($p\gamma$) reactions that produce charged pions. These pions subsequently decay into PeV-scale neutrinos \cite{Learned:2000sw}. UHE photons, by contrast, form through both leptonic and hadronic channels: relativistic electrons generate photons via synchrotron radiation or inverse Compton scattering, while hadronic processes (e.g., $pp$ or $p\gamma$ interactions) produce high-energy pions whose decay yields TeV–PeV photons~\cite{Cao:2023mig, Gill:2022erf}.

Multi-messenger observations of neutrinos and photons offer unparalleled insights into the physical conditions of cosmic accelerators, yet confirmed high-energy multi-messenger events remain scarce. Early examples include MeV-scale neutrino detections from SN 1987A~\cite{Kamiokande-II:1987idp} and the Sun~\cite{Bahcall:1987jc}, while more recently, the blazar TXS 0506+056 was associated with a ~290 TeV IceCube neutrino~\cite{IceCube:2018dnn} and enhanced 80–400 GeV gamma-ray emission (MAGIC)~\cite{MAGIC:2018sak}. These observations highlight the need to expand beyond traditional particle production frameworks: UHE neutrinos generated in dense regions can traverse to outer, optically thin layers of astrophysical environments, where they interact with nucleons to produce observable UHE photons—addressing the core limitation of photon escape in conventional models.

In this work, we aim to fill these gaps by modeling UHE photon production via neutrino–nucleon scattering in optically thin astrophysical regions. We consider incident (anti)neutrinos with energies spanning 1 TeV to 220 PeV (consistent with recent IceCube and KM3Net observations) and perform a systematic calculation of the resulting photon energy spectra. Our methodology includes three core steps: (i) computing neutrino–parton cross sections for both charged-current and neutral-current processes; (ii) integrating nucleon parton distribution functions (PDFs) and pion fragmentation functions to derive the neutrino–nucleon interaction cross section for 
$\pi^0$ production; and (iii) calculating the decay rate of $\pi^0$ (at given energies) into photons to obtain the photon energy probability distribution. Note that in this work we consider only UHE photons originating from $\pi^{0}$ decay, since for other hadrons the branching ratios of decay channels with photons are very small~\footnote{For instance, the branching ratio of $\pi^{\pm}\to \ell^{\pm}\nu\gamma$ is of order $\mathcal{O}(10^{-4})$, much smaller than the branching ratio of $\pi^{0}\to\gamma\gamma$ ($\simeq 98.8\%$)~\cite{ParticleDataGroup:2024cfk}.}.

In addition to the microscopic calculation, we further apply our results to a concrete astrophysical scenario by modeling the preburst emission of GRB~221009A, thereby connecting the underlying particle-level process to observable signals.

This paper is arranged as follows. In Sec.~\ref{Calculation}, we present the computational methodology in detail, 
including the calculations of scattering cross sections and decay rates. We present and discuss the energy distribution of UHE photons from neutrino-nucleon interactions in Sec.~\ref{Results}, apply the mechanism to the preburst of GRB 221009A in Sec.~\ref{sec:grb221009a}, and summarize our key conclusions in Sec.~\ref{Summary}. 

\section{Calculation}
\label{Calculation}
Recent multi-messenger observations, spanning UHE neutrinos and photons, have proliferated, yet a direct causal link between them remains unestablished. Therefore, quantifying the probability that a single UHE (anti)neutrino produces an observable TeV–PeV photon through charged and neutral current scattering on nucleons is essential for assessing astrophysical scenarios where the neutrino and photon generations are connected, or even causal. In this section, we develop the calculation and derive the neutrino-induced photon-production probability, showing that under plausible source conditions this probability can be substantial.

\subsection{Overview}\label{sec:overall}
High-energy photon production from (anti)neutrinos of energy $E_\nu$ proceeds in two steps: (i) $\nu/\bar\nu$ scatter on nucleons ($N=p,n$) producing partons that hadronize into $\pi^0$ with momentum $p_\pi$; (ii) the neutral pions decay via $\pi^0\!\to\!\gamma\gamma$ to photons of energy $E_\gamma$. The probability for producing a $\pi^0$ with momentum $p_\pi$ is proportional to the neutrino–nucleon cross section,
\begin{align}
	\label{eq:overall_cross_section}
	&\sigma_{\nu \to \pi^{0}}(p_{\pi},E_{\nu})= \\ \notag
	& \frac{1}{2}\sum_{N,i,i'}\int_{0}^{1}\mathrm{d}x\int_{0}^{1}\mathrm{d}z 
	\ f_{i}^{N}(x)\hat{\sigma}_{i,i'}\left(x,z\right)D^{\pi^0}_{i'}(z),
\end{align}
where $x$ is the parton momentum fraction in the nucleon, $z$ the momentum fraction carried by the hadronized $\pi^0$, $f_i^N(x)$ the parton distribution function (PDF), $D^{\pi^0}_{i'}(z)$ the fragmentation function (FF), and $\hat\sigma_{i\to i'}$ the parton-level cross section, which reads
\begin{align}
	\label{eq:parton_cross_section}
	\hat{\sigma}_{i,i'}=\frac{1}{2E_{\nu}2E_{i}|v_{\nu}-v_{i}|}\int \frac{\mathrm{d}^3p_{i'}}{(2\pi)^3 2E_{i'}}\frac{\mathrm{d}^3p_{l}}{(2\pi)^3 2E_{l}} \\ \notag
	\times|\mathcal{M}_{i,i'}|^2(2\pi)^4\delta^{(4)}\left(P_{\nu}+P_{i}-P_{l}-P_{i'} \right).
\end{align}
Here $p$, $P$, $E$, and $v$ denote a particle’s three-momentum, four-momentum, energy, and velocity, respectively. In Eq.~\eqref{eq:parton_cross_section}, $\mathcal{M}$ is the parton-level scattering amplitude; explicit expressions are given in Sec.~\ref{sec:Electroweak}.

The photons arise from neutral-pion decay, $\pi^0\!\to\!\gamma\gamma$. The conditional probability $P(E_\gamma|p_\pi)$ follows from the decay rate,
\begin{align}
	\label{eq:pion_decay_rate}
	\Gamma_{\pi^{0}\to \gamma}(E_{\gamma}, p_{\pi}) = \frac{1}{2}\frac{1}{2E_{\pi}}\int \frac{\mathrm{d}^3p_{\gamma}}{(2\pi)^32E_{\gamma}} \frac{\mathrm{d}^3p_{\gamma'}}{(2\pi)^32E_{\gamma'}} \\ \notag
	\times |\mathcal{M}_{\pi^0 \to \gamma + \gamma}|^2(2\pi)^4\delta^{(4)}\left(P_{\pi}-P_{\gamma}-P_{\gamma'} \right),
\end{align}
where the factor $1/2$ accounts for identical photons.

To obtain the probability that a photon with energy $E_{\gamma}$ is produced by the (anti)neutrino with energy $E_{\nu}$, denoted as $P(E_{\gamma}|E_{\nu})$, we should obtain the probability $P(p_{\pi}|E_{\nu})$, and $P(E_{\gamma}|p_{\pi})$. Then
\begin{equation}
	\label{eq:probability_formula_1}
	P(E_{\gamma}|E_{\nu})=\int_0^{\infty}  P(E_{\gamma}|p_{\pi})f(p_{\pi}|E_{\nu}) \ \mathrm{d}p_{\pi},
\end{equation}
where $f(p_{\pi}|E_{\nu})\equiv \mathrm{d}P(p_{\pi}|E_{\nu})/\mathrm{d}p_{\pi}$ is the probability density function. Since $P(p_{\pi}|E_{\nu}) \propto \sigma_{\nu \to \pi^{0}}$ and $P(E_{\gamma}|p_{\pi}) \propto \Gamma_{\pi^0 \to \gamma}$, the photon–energy density $f(E_{\gamma}|E_{\nu}) \equiv \mathrm{d}P(E_{\gamma}|E_{\nu})/\mathrm{d}E_{\gamma}$ follows as
\begin{equation}
	\label{eq:probability_formula_2}
	f(E_{\gamma}|E_{\nu})\propto \int_0^{\infty} \frac{\mathrm{d}\Gamma_{\pi^{0}\to \gamma}(E_{\gamma}, p_{\pi})}{\mathrm{d} E_{\gamma}} \frac{\mathrm{d}\sigma_{\nu \to \pi^{0}}(p_{\pi},E_{\nu})}{\mathrm{d}p_{\pi}} \mathrm{d}p_{\pi}.
\end{equation}
In next subsections, we will compute $\mathrm{d}\Gamma_{\pi^{0}\to \gamma}/ \mathrm{d} E_{\gamma}$ and $\mathrm{d}\sigma_{\nu \to \pi^{0}}/ \mathrm{d} p_{\pi}$, and then $f(E_{\gamma}|E_{\nu})$ can be solved.

\subsection{Pion Decay Part}\label{sec:decay}

\begin{figure}[t]
	\centering
	\includegraphics[width=0.95\linewidth]{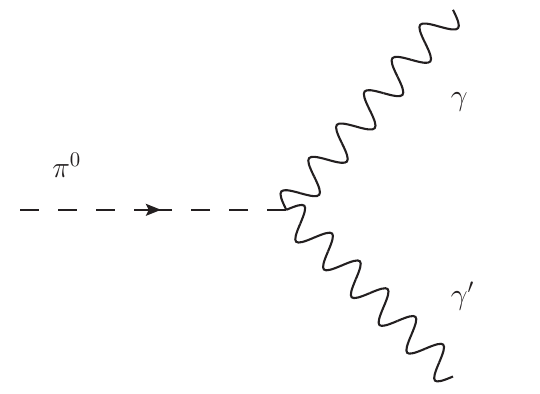}
	\caption{
		Feynman diagram for the neutral-pion decay.
	}
	\label{fig:Process00}
\end{figure}

In this section, we compute $\mathrm{d}\Gamma_{\pi^{0}\to \gamma}/ \mathrm{d} E_{\gamma}$ from the reaction $\pi^{0}\to\gamma\gamma$ (Fig.~\ref{fig:Process00}). The four-momenta are
\begin{align}
	\label{eq:four-momentum-one}
	P_{\pi}&=\left(\sqrt{p_{\pi}^2+m_{\pi}^2},0,0,p_{\pi} \right),\\ \notag
	P_{\gamma}&=\left(E_{\gamma}, E_{\gamma}\sin \theta \cos \phi, E_{\gamma}\sin \theta \sin \phi, E_{\gamma}\cos \theta\right),\\ \notag
	P_{\gamma'}&=\left(E_{\gamma'}, p_{\gamma',x},  p_{\gamma',y},  p_{\gamma',z}\right),
\end{align}
where $m_{\pi}$ is the neutral pion mass. Based on the energy-momentum conservation $P_{\pi}=P_{\gamma}+P_{\gamma'}$, we have
\begin{align}
	\label{eq:inner_product_pion_dacay_one}
	\left(P_{\gamma}\cdot P_{\gamma}\right)=\left(P_{\gamma'}\cdot P_{\gamma'}\right)=0,\ \ \ 
	\left(P_{\gamma}\cdot P_{\gamma'}\right)=\frac{m_{\pi}^2}{2}.
\end{align}
Additionally, $E_{\gamma}$ can be solved as
\begin{equation}
	\label{eq:E_gamma_expression}
	E_{\gamma}=\frac{m_{\pi}^2}{2\left(\sqrt{p_{\pi}^2+m_{\pi}^2}-p_{\pi}\cos \theta\right)}.
\end{equation}
Then, for $\theta$ varies in the region $[0, \pi]$, Eq.~\eqref{eq:E_gamma_expression} indicates that $E_{\gamma}$ is constrained in the range
\begin{equation}
	\label{eq:E_gamma_range}
	\frac{\sqrt{p^2_{\pi}+m_{\pi}^2}-p_{\pi}}{2} \leq E_{\gamma} \leq \frac{\sqrt{p^2_{\pi}+m_{\pi}^2}+p_{\pi}}{2}.
\end{equation}

After finishing the kinematical analysis, then we focus on the dynamical issue. We solve the amplitude square $|\mathcal{M}_{\pi^{0}\to \gamma \gamma}|^2$ shown in Eq.~\eqref{eq:pion_decay_rate}, which comes from the following effective Lagrangian 
\begin{equation}
	\label{eq:Lagrangian_pion_decay}
	\mathcal{L}=-\frac{\alpha}{8\pi f_{\pi}}\epsilon^{\mu \nu \alpha \beta} \pi^{0} F_{\mu\nu}F_{\alpha \beta},
\end{equation}
where $\alpha$ is fine structure constant, $f_{\pi}$ is the pion decay constant, $\epsilon^{\mu \nu \alpha \beta}$ is the Levi-Civita symbol and $F_{\mu \nu}=\partial_{\mu}A_{\nu}-\partial_{\nu}A_{\mu}$ is the electromagnetic tensor with electromagnetic potential $A_{\mu}$. Then the amplitude is
\begin{equation}
	\label{eq:scatter_amplitude_pion_decay}
	\mathcal{M}_{\pi^0 \to \gamma + \gamma}=i\frac{\alpha}{\pi f_{\pi}} \epsilon_{\mu\nu\alpha\beta}P^{\mu}_{\gamma}\epsilon_{k}^{*\nu}P^{\alpha}_{\gamma'}\epsilon_{k'}^{*\beta},
\end{equation}
where $\epsilon_{k}^{*\nu}$ and $\epsilon_{k'}^{*\beta}$ are the photon polarization vectors. Now the amplitude square is solved as
\begin{align}
	\label{eq:amplitude_square_pion_decay}
	|\mathcal{M}_{\pi^0 \to \gamma \gamma}|^2
	=\frac{\alpha^2 m_{\pi}^4}{2\pi^2 f_{\pi}^2},
\end{align}
where the relationship $\sum \epsilon^{\mu}_{k}\epsilon^{*\nu}_{k}=-g^{\mu\nu}$, $\epsilon_{\mu\nu\alpha\beta}\epsilon_{\mu'\ \alpha'}^{\ \ \nu\ \beta}=2(g_{\mu \alpha'}g_{\alpha \mu'}-g_{\mu \mu'}g_{\alpha \alpha'})$ have been used.

After deriving the amplitude square Eq.~\eqref{eq:amplitude_square_pion_decay}, now we can go further to calculate $\mathrm{d}\Gamma_{\pi^{0}\to \gamma}/ \mathrm{d} E_{\gamma}$. Through Eq.~\eqref{eq:pion_decay_rate}, we have
\begin{align}
	\label{eq:pion_decay_rata_derivation_two}
	&\frac{\mathrm{d}\Gamma_{\pi^0 \to \gamma}\left(E_{\gamma},p_{\pi}\right)}{\mathrm{d}E_{\gamma}}= \\ \notag
    &\frac{\alpha^2m_{\pi}^4}{64\pi^3 f_{\pi}^2\sqrt{p_{\pi}^2+m_{\pi}^2}}\int_{0}^{\pi} \frac{E_{\gamma} \sin \theta \mathrm{d}\theta}{\sqrt{E_{\gamma}^2+p_{\pi}^2-2p_{\pi}E_{\gamma}\cos \theta}} \\ \notag
	 &\times \delta\left(\sqrt{p_{\pi}^2+m_{\pi}^2} - \sqrt{E_{\gamma}^2+p_{\pi}^2-2p_{\pi}E_{\gamma}\cos \theta}-E_{\gamma}\right).
\end{align}
For the $\theta$–integration in Eq.~\eqref{eq:pion_decay_rata_derivation_two}, the kinematics imply that if $E_{\gamma}$ lies outside Eq.~\eqref{eq:E_gamma_range}, the equation $\sqrt{p_{\pi}^2+m_{\pi}^2} - \sqrt{E_{\gamma}^2+p_{\pi}^2-2p_{\pi}E_{\gamma}\cos \theta}-E_{\gamma}=0$ has no solution, so $\mathrm{d}\Gamma_{\pi^{0}\to\gamma}/\mathrm{d}E_{\gamma}=0$. 
If $E_{\gamma}$ satisfies Eq.~\eqref{eq:E_gamma_range}, we have
\begin{equation}
	\label{eq:cos_thetha_one}
	\cos \theta = \frac{2E_{\gamma}\sqrt{p_{\pi}^2+m_{\pi}^2}-m_{\pi}^2}{2p_{\pi}E_{\gamma}}.
\end{equation}
Then inserting this into Eq.~\eqref{eq:pion_decay_rata_derivation_two}, the non-zero $\mathrm{d}\Gamma_{\pi^{0}\to \gamma}/\mathrm{d}E_{\gamma}$ is obtained.
Finally, the result is
\begin{align}
	\label{eq:final_result_one}
	&\frac{\mathrm{d}\Gamma_{\pi^0 \to \gamma}\left(E_{\gamma},p_{\pi}\right)}{\mathrm{d}E_{\gamma}}= \\ \notag
	&\begin{cases}
		\frac{\alpha^2m_{\pi}^4}{64\pi^3 f_{\pi}^2p_{\pi}\sqrt{p_{\pi}^2+m_{\pi}^2}}, &  \frac{\sqrt{p^2_{\pi}+m_{\pi}^2}-p_{\pi}}{2} \leq E_{\gamma} \leq \frac{\sqrt{p^2_{\pi}+m_{\pi}^2}+p_{\pi}}{2},\\
		0, & \mathrm{otherwise}.
	\end{cases}
\end{align}
With $\mathrm{d}\Gamma_{\pi^{0}\to \gamma}/\mathrm{d}E_{\gamma}$ in hand, we next derive $\mathrm{d}\sigma_{\nu \to \pi^{0}}/\mathrm{d}p_{\pi}$ in the following subsections.

\subsection{Phase Space Integration}\label{sec:integral}
After obtaining $\mathrm{d}\Gamma_{\pi^{0}\to \gamma}/ \mathrm{d} E_{\gamma}$, we now turn to $\mathrm{d}\sigma_{\nu \to \pi^{0}}/ \mathrm{d} p_{\pi}$. Our goal in this section is to solve the phase space integration in Eq.~\eqref{eq:parton_cross_section}. We begin with the kinematics. Here, for convenience we work in the centre-of-mass frame of the (anti)neutrino and nucleon (denoted as $\nu N$-frame in the following). At ultra-high energies the invariant mass is large, so we neglect the masses of all external particles in the $\nu N$-frame. The four-momenta are given as
\begin{align}
	\label{eq:four-momentum-two}
	&P_{\nu}=\left(E_{\nu}, 0, 0, E_{\nu}\right), \ \ \ P_{N}=\left(E_{\nu},0,0, -E_{\nu}\right),\\ \notag
	&P_{i} = \left(xE_{\nu},0,0,-xE_{\nu}\right), \ \ \ P_{l}=\left(E_{l}, p_{l,x}, p_{l,y}, p_{l,z}\right), \\ \notag
	&P_{i'}=\left(\frac{p_{\pi}}{z},\frac{p_{\pi}}{z}\sin \theta \cos \phi, \frac{p_{\pi}}{z} \sin \theta \sin \phi, \frac{p_{\pi}}{z} \cos \theta \right),
\end{align}
where the initial parton carries momentum fraction $x$ of the nucleon, and the final-state parton transfers a fraction $z$ to the $\pi^{0}$. From momentum conservation $P_{\nu}+P_{i}=P_{l}+P_{i'}$, we obtain the relations
\begin{align}
	\label{inner_product_formula_one}
	& P_{\nu} \cdot P_{i} = 2xE_{\nu}^2, \\
	\label{inner_product_formula_two}
	& P_{\nu} \cdot P_{l} = 2xE_{\nu}^2-\frac{p_{\pi}E_{\nu}}{z}\left(1-\cos \theta\right).
\end{align}
Additionally, $p_{\pi}$ can be solved through the relation $E_{l}^2\simeq \sqrt{p_{l,x}^2+p_{l,y}^2+p_{l,z}^2}$, with
\begin{equation}
	\label{eq:pion_momentum_expression}
	p_{\pi}=\frac{2xzE_{\nu}}{(1+x)-(1-x)\cos \theta}.
\end{equation}
which restricts $p_{\pi}$ to the range
\begin{equation}
	\label{eq:pion_momentum_range}
	xzE_{\nu} \leq p_{\pi} \leq zE_{\nu}.
\end{equation}
With these kinematics and Eq.~\eqref{eq:parton_cross_section}, we have
\begin{align}
	\label{eq:derivation_cross_section_two}
	&\frac{\mathrm{d}\hat{\sigma}_{i,i'}(p_{\pi},E_{\nu})}{\mathrm{d}p_{\pi}} = \frac{p_{\pi}}{64\pi xz^2E_{\nu}^2} \int_{0}^{\pi} \mathrm{d}\theta  \sin \theta \ |\mathcal{M}_{i,i'}|^2 \\ \notag
	& \times \frac{1}{\sqrt{(p_{\pi}/z)^2+(1-x)^2 E_{\nu}^2 - 2(1-x)E_{\nu}(p_{\pi}/z)\cos \theta}} \\ \notag
	&\times \delta \bigg ((1+x)E_{\nu}-\frac{p_{\pi}}{z}  \\ \notag
	& -\sqrt{\left(\frac{p_{\pi}}{z}\right)^2+(1-x)^2 E_{\nu}^2 - 2(1-x)E_{\nu}\left(\frac{p_{\pi}}{z}\right)\cos \theta}\ \bigg ).
\end{align}
As in the decay case, the $\theta$–integration is controlled by kinematics. If $p_{\pi}$ lies outside the allowed range \eqref{eq:pion_momentum_range}, the argument of the delta function in Eq.~\eqref{eq:derivation_cross_section_two} has no solution for $\theta\in[0,\pi]$, giving $\mathrm{d}\hat{\sigma}_{i,i'}/\mathrm{d}p_{\pi}=0$. Within the allowed region, $\cos\theta$ is fixed by
\begin{equation}
	\label{eq:cos_thetha_two}
	\cos \theta = \frac{(1+x)p_{\pi}-2xzE_{\nu}}{(1-x)p_{\pi}}.
\end{equation}
leading directly to the non-zero result.
Now, we obtain the final result of $\mathrm{d}\hat{\sigma}_{i,i'}/\mathrm{d}p_{\pi}$, which reads
\begin{align}
	\label{eq:final_result_two}
	\frac{\mathrm{d}\hat{\sigma}_{i,i'}(p_{\pi},E_{\nu})}{\mathrm{d}p_{\pi}}=
	\begin{cases}
		\frac{|\mathcal{M}_{i,i'}|^2}{64\pi x(1-x)zE_{\nu}^3}, &  xzE_{\nu} \leq p_{\pi} \leq zE_{\nu},\\
		0, & \mathrm{otherwise}.
	\end{cases}
\end{align}
With this in hand, we then compute $|\mathcal{M}_{i,i'}|^{2}$ in Sec.~\ref{sec:Electroweak} to obtain the final $\mathrm{d}\hat{\sigma}_{i,i'}/\mathrm{d}p_{\pi}$.

\subsection{Electroweak Part}\label{sec:Electroweak}
\begin{figure*}[t]
	\centering
	\includegraphics[width=0.45\linewidth]{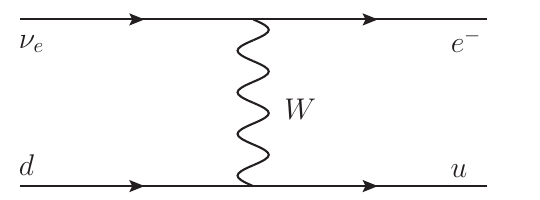}
	\includegraphics[width=0.45\linewidth]{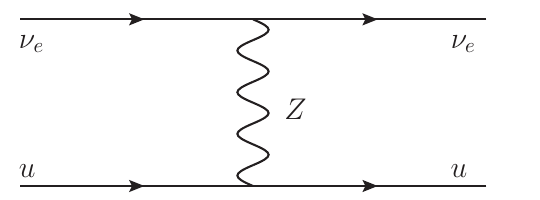}
	\caption{
		Representative electroweak diagrams for neutrino–parton scattering: 
		(left) charged-current process mediated by a $W$ boson; 
		(right) neutral-current process mediated by a $Z$ boson.
	}
	\label{fig:Feynman}
\end{figure*}

In this section we compute the amplitude squares $|\mathcal{M}_{i,i'}|^{2}$ that enter Eq.~\ref{eq:final_result_two}. (Anti)neutrino–parton scattering proceeds through two types of electroweak interactions: charged-current (CC) processes mediated by $W$ bosons and neutral-current (NC) processes mediated by $Z$ bosons. Representative Feynman diagrams for each class are shown in Fig.~\ref{fig:Feynman}.

For the CC process in the left panel of Fig.~\ref{fig:Feynman}, the amplitude square is
\begin{align}
	\label{eq:process_one}
	&|\mathcal{M}_{d,u}|^2=|V_{ud}|^2\frac{e^4}{\sin^4 \theta_w} \\ \notag
	&\times \frac{P_{\nu} \cdot P_{i}\left(m_d^2-m_e^2-m_u^2+2P_{\nu} \cdot P_{i}\right)}{\left(m_{W}^2-m_e^2+2P_{\nu} \cdot P_{l} \right)^2}.
\end{align}
Here $e$ is the electric charge, $\theta_w$ is the weak mixing angle ($\sin^2\theta_w\simeq 0.23$), and $V_{ud}$ is a CKM matrix element. 

For the NC process in the right panel of Fig.~\ref{fig:Feynman}, we obtain
\begin{align}
	\label{eq:process_two}
	&|\mathcal{M}_{u,u}|^2=\frac{e^4}{\cos^4 \theta_w} \frac{1}{18\left(m_{Z}^2+2P_{\nu} \cdot P_{l}\right)^2} \\ \notag
	&\times \bigg[4 P_{\nu} \cdot P_{l} \left(m_{u}^2 \left(3 \csc^2 \theta_w -4\right)-8 P_{\nu} \cdot P_{i} \right)  \\ \notag
	&+ \left(9 \csc^4 \theta_w -24 \csc^2 \theta_w + 32\right) \left(P_{\nu} \cdot P_{i}\right)^2 + 16 \left(P_{\nu} \cdot P_{l}\right)^2\bigg].
\end{align}
Equations~\eqref{eq:process_one} and \eqref{eq:process_two} correspond to two representative channels. A complete list of all relevant processes and their amplitude squares is given in Appendix~\ref{appendix:amplitude_square}.

To match the kinematic setup used in Sec.~\ref{sec:integral}, we now evaluate these amplitudes in the $\nu N$-frame, where particle masses can be neglected. Substituting Eqs.~\eqref{inner_product_formula_one}, \eqref{inner_product_formula_two}, and \eqref{eq:cos_thetha_two} into the amplitude squares, yields
\begin{align}
	\label{eq:process_one_CoM}
	&|\mathcal{M}_{d,u}|^2= \\ \notag
	&\frac{8e^4 E_{\nu}^4 |V_{ud}|^2 (x-1)^2 x^2 z^2}{\sin^4 \theta_w \left[4x E_{\nu}  \left(xzE_{\nu}  -p_{\pi}\right)+z(x-1) m_{W}^2\right]^2}
\end{align}
\begin{align}
	\label{eq:process_two_CoM}
	&|\mathcal{M}_{u,u}|^2=\\ \notag 
	&\frac{2e^4 x^2 E_{\nu}^2}{9 \cos^4 \theta_w \left[4 z x^2 E_{\nu}^2 -4 x E_{\nu} p_{\pi} + z (x-1)m_{Z}^2\right]^2} \\ \notag
	&\times \bigg[ 16 \left(E_{\nu}^2z^2 (x^2 - 2x +2) - 2 z E_{\nu}p_{\pi} +p_{\pi}^2\right)  \\ \notag
	&+3z^2(x-1)^2 E_{\nu}^2 \csc^4 \theta_w \left(8 \cos^2 \theta_w -5\right)  \bigg].
\end{align}
Finally, substituting \eqref{eq:process_one_CoM} and \eqref{eq:process_two_CoM} into Eq.~\ref{eq:final_result_two} gives the corresponding contributions to $\mathrm{d}\hat{\sigma}_{i,i'}/\mathrm{d}p_{\pi}$.

\subsection{PDF and  Fragmentation Function}\label{sec:PDF_and_FF}
In this section we briefly summarize the parton distribution functions (PDFs) and fragmentation functions (FFs) that enter the total cross section $\sigma_{\nu\to\pi^0}$ in Eq.~\eqref{eq:overall_cross_section}.

PDFs describe the interior structure of hadrons in deep-inelastic scattering: they give the probability of finding a parton carrying a momentum fraction $x$ of the parent hadron. Because PDFs are non-perturbative, they must be extracted from data rather than computed directly in perturbative QCD. In our calculation, the relevant hadrons are protons and neutrons, and we include the quark flavors $u$, $\bar u$, $d$, $\bar d$, $s$, and $\bar s$. For numerical input we use the PDF sets of Refs.~\cite{Schmidt:2015zda,Xie:2023qbn}; a general review can be found in Ref.~\cite{Lorce:2025aqp}.

Fragmentation functions describe the hadronization of an energetic parton into final-state hadrons. They encode the probability that a parton produces a hadron carrying a momentum fraction $z$ of the original parton. Like PDFs, FFs are non-perturbative quantities extracted from experiment, with more details discussed in Refs.~\cite{deFlorian:2007aj,deFlorian:2007ekg,deFlorian:2014xna,Peng:2023mxp,mcwy-b221}. In this work we adopt the parametrization of Ref.~\cite{deFlorian:2014xna}, where the FF for producing a $\pi^{+}$ is modeled by 
\begin{align}
	\label{eq:FF_fitting}
	&D_{i}^{\pi^{+}}(z)=\\ \notag &\frac{N_{i}z^{\alpha_{i}}(1-z)^{\beta_{i}}[1+\gamma_{i}(1-z)^{\delta_i}]}{B[2+\alpha_{i},\beta_{i}+1]+\gamma_{i}B[2+\alpha_{i},\beta_{i}+\delta_{i}+1]},
\end{align}
with $B[a,b]$ the Euler Beta function. The FFs for $\pi^{-}$ follow from charge conjugation, e.g.\ $D_{u}^{\pi^{-}}(z)=D_{\bar u}^{\pi^{+}}(z)$, and the neutral-pion FF is obtained through $D^{\pi^0}_{i}(z)=[D^{\pi^{+}}_{i}(z)+D^{\pi^{-}}_{i}(z)]/2$.

The fitted parameters for Eq.~\eqref{eq:FF_fitting} are listed in Table~\ref{tab:parameter}. Additional relations used in our calculation include~\cite{deFlorian:2014xna}
\begin{align}
	\label{eq:FF_property}
	&D_{\overline{u}}^{\pi^{+}}(z)=D_{d}^{\pi^{+}}, \ \ D^{\pi^{+}}_{s}(z)=D^{\pi^{+}}_{\overline{s}}(z), \\ \notag
	&D^{\pi^{+}}_{i+i'}(z) =D^{\pi^{+}}_{i}(z)+D^{\pi^{+}}_{i'}(z).
\end{align}
which, together with Table~\ref{tab:parameter}, allow us to obtain all parton FFs needed for $\pi^0$ production.
\begin{table}[ht]
	\centering
	\renewcommand{\arraystretch}{1.35} 
	\setlength{\tabcolsep}{4pt}
	\scalebox{1.2}{
		\begin{tabular}{*{12}{c}}
			\hline
			\hline
			Parton $i$  & $N_i$ & $\alpha_{i}$ & $\beta_{i}$ & $\gamma_{i}$ & $\delta_{i}$ \\
			\hline
			$u+\overline{u}$ & 0.387 & -0.388  & 0.910 & 7.15 & 3.96 \\
			\hline
			$d+\overline{d}$ & 0.388 & -0.388  & 0.910 & 7.15 & 3.96 \\
			\hline
			$\overline{u}=d$ & 0.105 & 1.649 & 3.286 & 49.95 &  8.67 \\
			\hline
			$s+\overline{s}$ & 0.273 & 1.449 & 3.286 & 49.95 &  8.67 \\
			\hline
			\hline
	\end{tabular}}
	\caption{Fitted parameters $(N_i,\alpha_i,\beta_i,\gamma_i,\delta_i)$ for the fragmentation function in Eq.~\eqref{eq:FF_fitting}, taken from Ref.~\cite{deFlorian:2014xna}. These parameters determine the $\pi^{0}$ FFs follow from charge conjugation, Eq.~\eqref{eq:FF_property} and $D^{\pi^0}_{i}(z)=[D^{\pi^{+}}_{i}(z)+D^{\pi^{-}}_{i}(z)]/2$ as described in the text.}
	\label{tab:parameter}
\end{table}

\subsection{Final Expression}\label{sec:final_result}

With all ingredients prepared, the probability density for producing a photon from an incident (anti)neutrino is given by
\begin{align}
	\label{eq:probability_formula_3}
	&f(E^{\rm lab}_{\gamma}|E^{\rm lab}_{\nu})\propto \\ \notag &\int_0^{\infty} \frac{\mathrm{d}\Gamma_{\pi^{0}\to \gamma}(E^{\nu N}_{\gamma}, p^{\nu N}_{\pi})}{\mathrm{d} E^{\nu N}_{\gamma}} \frac{\mathrm{d}\sigma_{\nu \to \pi^{0}}(p^{\nu N}_{\pi},E^{\nu N}_{\nu})}{\mathrm{d}p'_{\pi}} \mathrm{d}p^{\nu N}_{\pi},
\end{align}
where the ingredients are
\begin{align}
	\label{eq:probability_formula_4}
	&\frac{\mathrm{d}\Gamma_{\pi^0 \to \gamma}\left(E_{\gamma},p_{\pi}\right)}{\mathrm{d}E_{\gamma}}= \\ \notag
	&\begin{cases}
		\frac{\alpha^2m_{\pi}^4}{64\pi^3 f_{\pi}^2p_{\pi}\sqrt{p_{\pi}^2+m_{\pi}^2}}, &  \frac{\sqrt{p^2_{\pi}+m_{\pi}^2}-p_{\pi}}{2} \leq E_{\gamma} \leq \frac{\sqrt{p^2_{\pi}+m_{\pi}^2}+p_{\pi}}{2},\\
		0, & \mathrm{otherwise}.
	\end{cases}
\end{align}
\begin{align}
	\label{eq:probability_formula_5}
	&\frac{\mathrm{d}\sigma_{\nu \to \pi^{0}}(p_{\pi},E_{\nu})}{\mathrm{d}p_{\pi}}= \\ \notag
	&\frac{1}{2}\sum_{N,i,i'}\int_{0}^{1}\mathrm{d}x\int_{0}^{1}\mathrm{d}z 
	\ f_{i}^{N}(x)\frac{\mathrm{d}\hat{\sigma}_{i,i'}\left(p_{\pi},E_{v},x,z\right)}{\mathrm{d}p_{\pi}}D^{\pi^0}_{i'}(z),
\end{align}
\begin{align}
	\label{eq:probability_formula_6}
	\frac{\mathrm{d}\hat{\sigma}_{i,i'}(p_{\pi},E_{\nu},x,z)}{\mathrm{d}p_{\pi}}=
	\begin{cases}
		\frac{|\mathcal{M}_{i,i'}|^2}{64\pi x(1-x)zE_{\nu}^3}, &  xzE_{\nu} \leq p_{\pi} \leq zE_{\nu},\\
		0, & \mathrm{otherwise}.
	\end{cases}
\end{align}
The required inputs: PDFs $f_i^N(x)$, FFs $D_{i}^{\pi^{0}}(z)$, and amplitudes $|\mathcal{M}_{i,i'}|^{2}$, are provided in Sec.~\ref{sec:PDF_and_FF}, Sec.~\ref{sec:Electroweak}, Appendix~\ref{appendix:amplitude_square}, and Refs.~\cite{deFlorian:2014xna,Schmidt:2015zda,Xie:2023qbn}.

 Note that in Eq.~\eqref{eq:probability_formula_3} we use superscripts to distinguish quantities in the $\nu N$-frame and laboratory frame. Since Eqs.~\eqref{eq:probability_formula_4}–\eqref{eq:probability_formula_6} are constructed in the $\nu N$-frame, while the observable probability density is defined in the laboratory frame, a Lorentz transformation is required to relate $(E^{\rm lab}_{\nu},E^{\rm lab}_{\gamma})$ to $(E^{\nu N}_{\nu},E^{\nu N}_{\gamma})$.

In the laboratory frame, nucleons are effectively at rest: their thermal energy is $\mathcal{O}(\mathrm{MeV})$ in optically thin astrophysical regions~\cite{suzuki2024neutrinos}, far below the nucleon mass $m_{N}\simeq 1~\mathrm{GeV}$. The relative velocity between the two frames is therefore
\begin{equation}
	\label{eq:beta}
	v = \frac{E^{\rm lab}_{\nu}}{E^{\rm lab}_{\nu}+m_{N}}.
\end{equation}

Additionally, Refs.~\cite{Dar:1999ie,Nakar:2015tma} demonstrate that UHE photons and neutrinos originating from the same source propagate nearly collinearly. Thus both energies transform in the same way:
\begin{equation}
	\label{eq:transform}
	E^{\rm lab}_{\nu / \gamma} \to E^{\nu N}_{\nu / \gamma} = E^{\rm lab}_{\nu / \gamma}  \sqrt{\frac{1-v}{1+v}}.
\end{equation}

To summarize, in this section we have assembled all ingredients needed to compute the conditional probability density $f(E^{\rm lab}_{\gamma}|E^{\rm lab}_{\nu})$: the $\pi^{0}\!\to\!\gamma\gamma$ decay rate, the (anti)neutrino induced $\pi^{0}$ production cross section, the underlying electroweak matrix elements, and the nonperturbative PDFs and FFs. Together with the Lorentz transformation between the $\nu N$-frame and the laboratory frame,  Eqs.~\eqref{eq:probability_formula_3}–\eqref{eq:probability_formula_6} fully determine the photon production probability from an incident (anti)neutrino of energy $E_{\nu}$. In the next section we evaluate these expressions numerically and explore the resulting $E_{\gamma}$ distributions and their phenomenological implications for UHE astrophysics.

\section{Results}
\label{Results}
\begin{figure*}[htbp]
    \centering
    \begin{minipage}{0.48\textwidth}
        \centering
        \includegraphics[width=\linewidth]{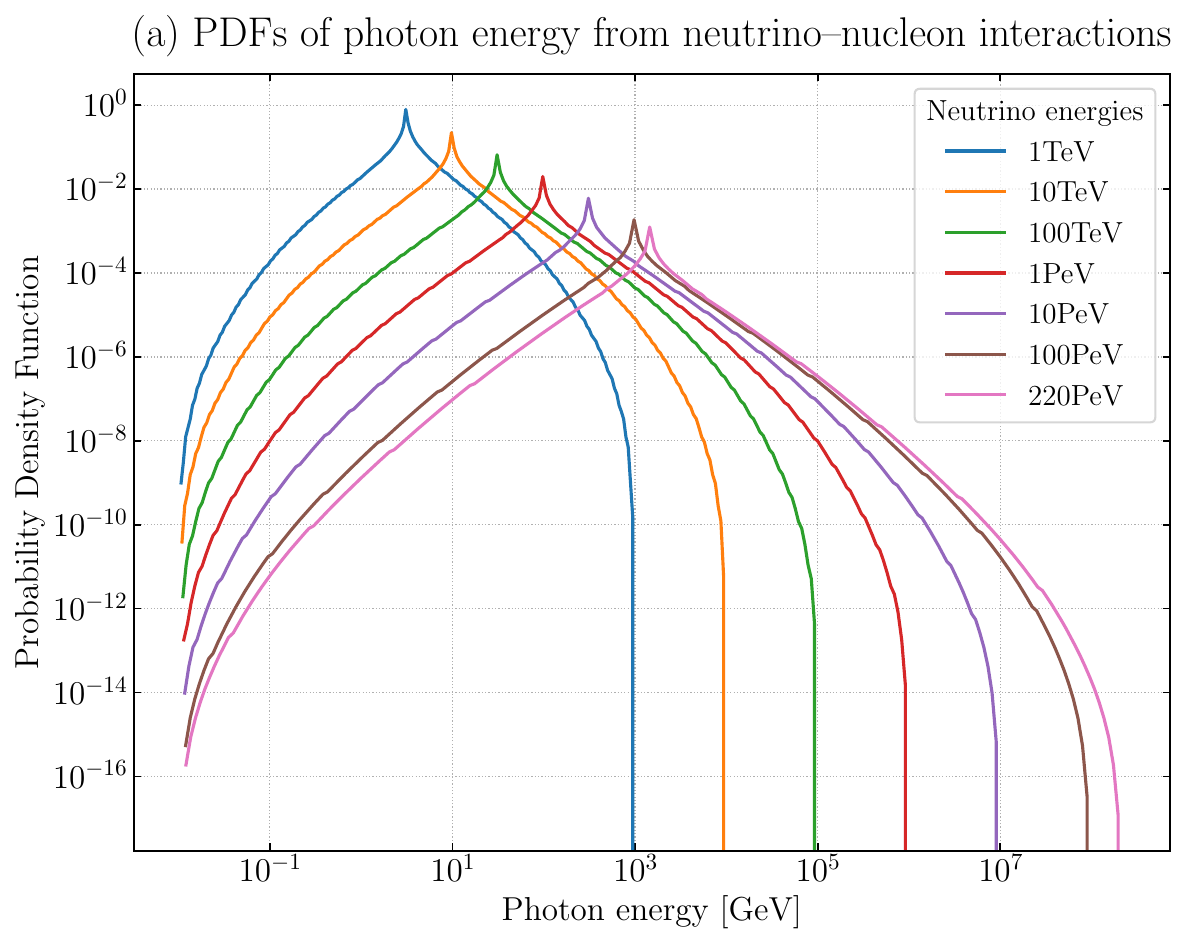}
    \end{minipage}\hfill
    \begin{minipage}{0.48\textwidth}
        \centering
        \includegraphics[width=\linewidth]{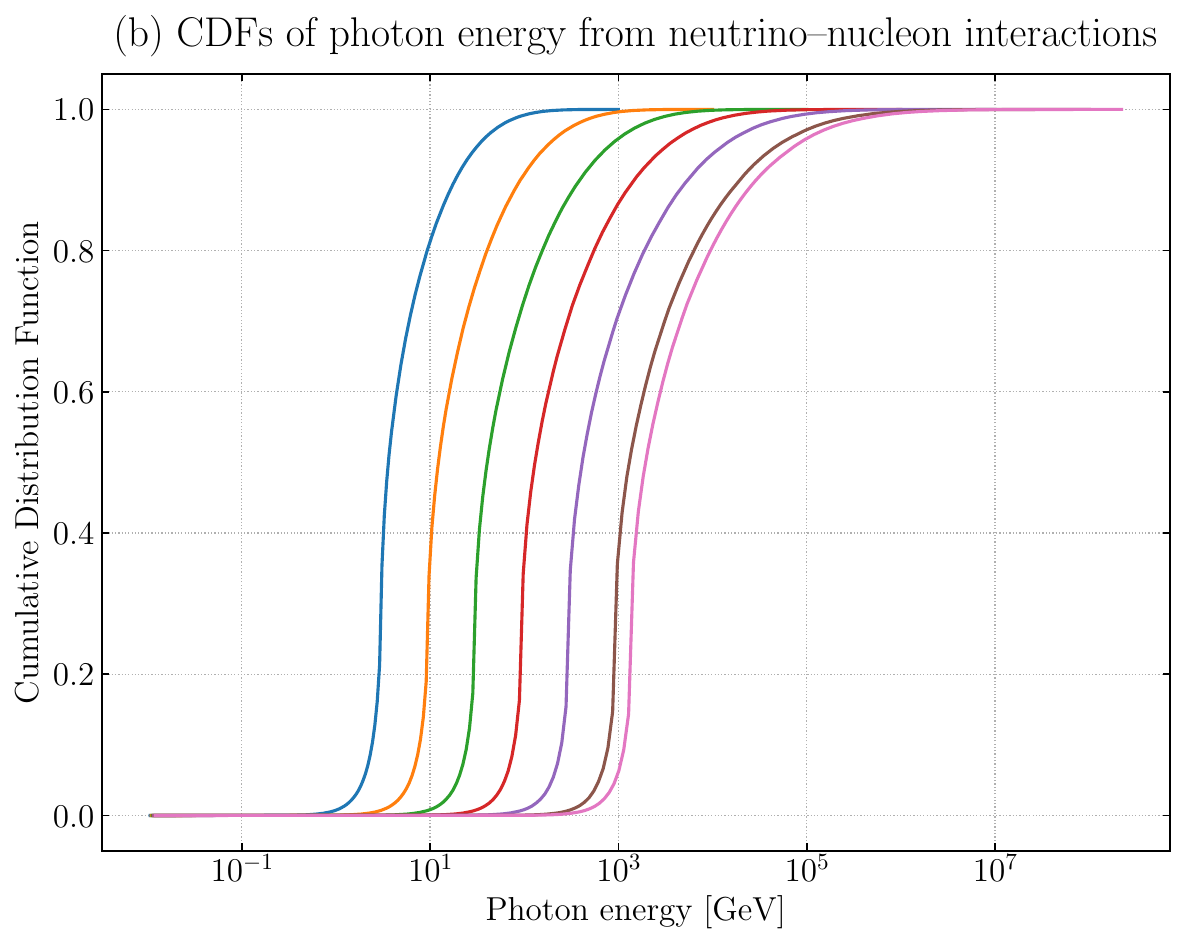}
    \end{minipage}
    
    \begin{minipage}{0.48\textwidth}
        \centering
        \includegraphics[width=\linewidth]{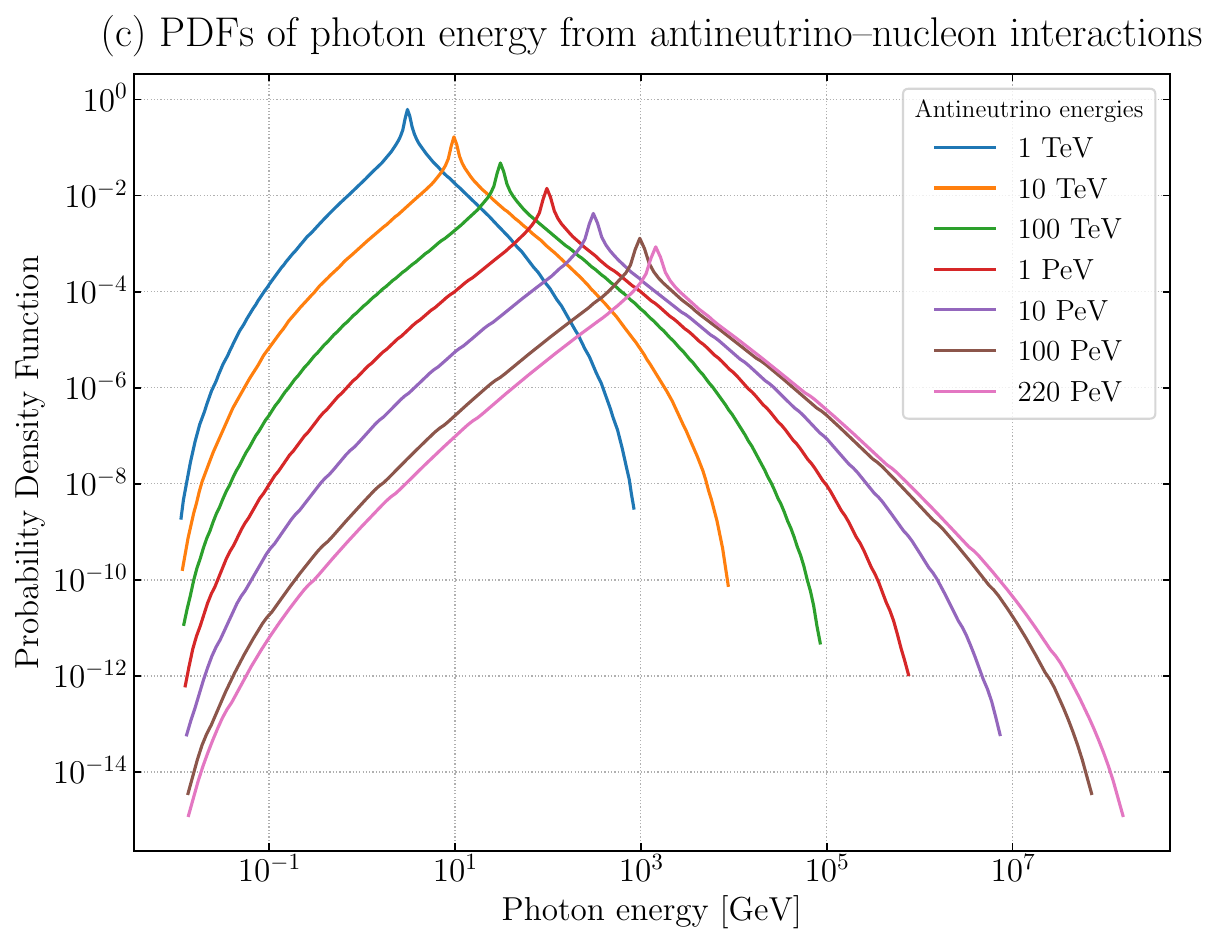}
    \end{minipage}\hfill
    \begin{minipage}{0.48\textwidth}
        \centering
        \includegraphics[width=\linewidth]{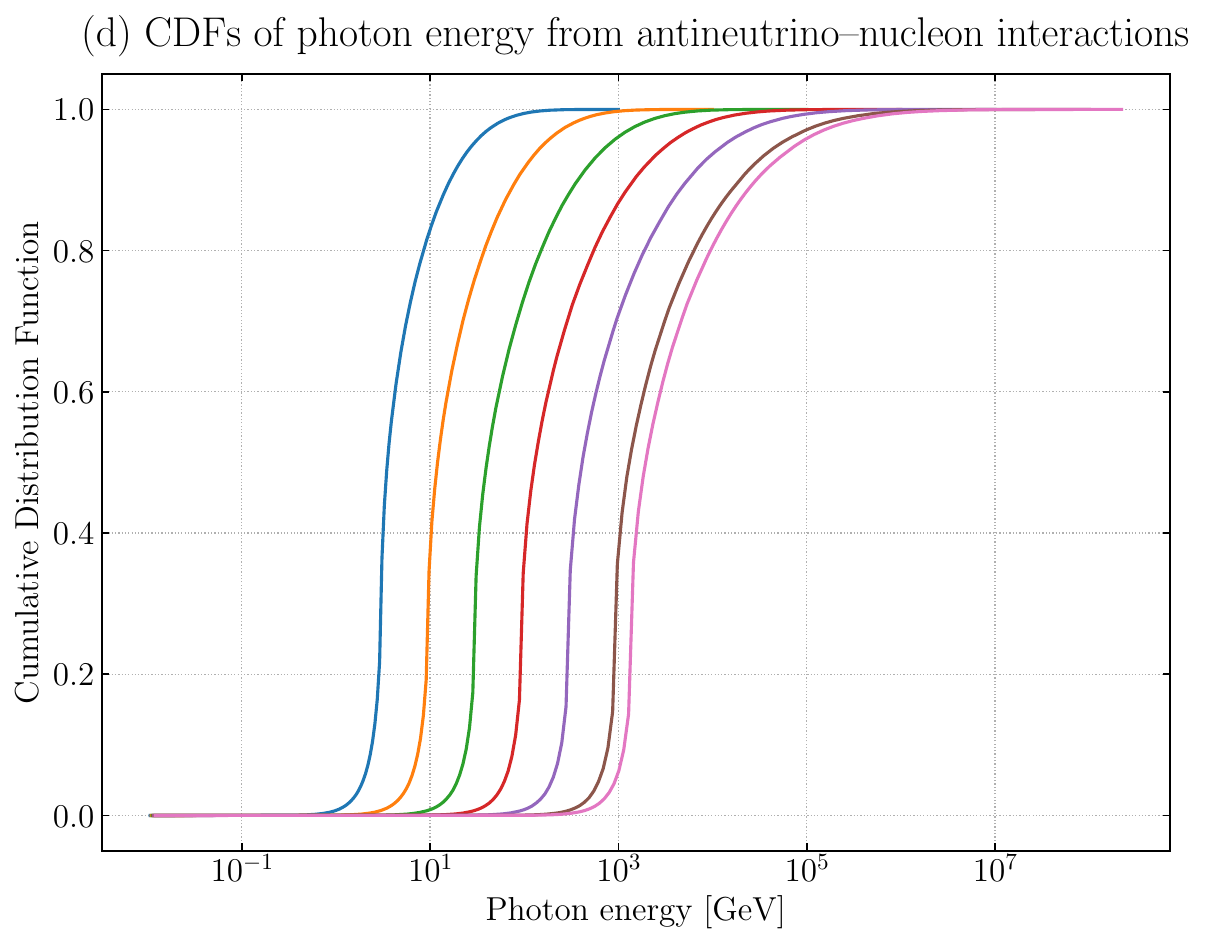}
    \end{minipage}
    
    \caption{Photon energy distributions generated by neutrino–nucleon and antineutrino–nucleon scattering.
    The upper panels show the probability density (left) and cumulative distribution (right) for neutrinos, while the lower panels present the corresponding distributions for antineutrinos. The results demonstrate that TeV–PeV photons are produced with significant probability in both cases.}
    \label{fig:results}
\end{figure*}

With the derivations in Sec.~\ref{Calculation}, we can now evaluate the probability distribution for photon production induced by UHE neutrinos and antineutrinos. The results are shown in Fig.~\ref{fig:results}. The upper (lower) panels correspond to neutrinos (antineutrinos), while the left and right panels display the probability density functions and cumulative distribution functions, respectively. As illustrated in Fig.~\ref{fig:results}, high-energy (anti)neutrino-nucleon scattering has a sizable probability of producing energetic photons.

Figure~\ref{fig:results} exhibits two notable features of the photon distribution:
\begin{itemize}
\item The probability density is symmetric in $\log E_{\gamma}$.

\item The lower bound on $E_{\gamma}$ is nearly independent of the incident (anti)neutrino energy $E_{\nu}$.
\end{itemize}
We now explain these features analytically. For clarity, we use superscripts to distinguish quantities defined in different frames.

To explain the symmetry of the photon probability density function, we focus on Eqs.~\eqref{eq:probability_formula_3}–\eqref{eq:probability_formula_6}. These show that the dependence of $f(E^{\rm lab}_{\gamma}|E^{\rm lab}_{\nu})$ on $E^{\rm lab}_{\gamma}$ (or equivalently on $E^{\nu N}_{\gamma}$) arises solely through the kinematic constraint $(\sqrt{(p^{\nu N}_{\pi})^2+m_{\pi}^2}-p^{\nu N}_{\pi})/2 \leq E^{\nu N}_{\gamma} \leq (\sqrt{(p^{\nu N}_{\pi})^2+m_{\pi}^2}+p^{\nu N}_{\pi})/2$. In other words, $E^{\nu N}_{\gamma}$ only alters the integration domain in Eq.~\eqref{eq:probability_formula_3}, but not the integrand. This constraint implies that the integral in Eq.~\eqref{eq:probability_formula_3} has a lower limit $p^{\nu N}_{\rm low}$, which depends on $E^{\nu N}_{\gamma}$ and is given by:
\begin{equation}
    \label{eq:lower_bound}
    p^{\nu N}_{\rm low}=\frac{|m_{\pi}^2-4(E^{\nu N}_{\gamma})^2|}{4E^{\nu N}_{\gamma}}.
\end{equation}
Crucially, both $E^{\nu N}_{\gamma}$ and $m_{\pi}^2 / (4 E^{\nu N}_{\gamma})$ lead to the same $p^{\nu N}_{\rm low}$. Therefore, the integrand and integration region in Eq.~\eqref{eq:probability_formula_3} are identical for these two energies. Giving
\begin{equation}
f(E^{\nu N}_{\gamma}|E^{\nu N}_{\nu})=f\left(\frac{m_{\pi}^2}{4 E^{\nu N}_{\gamma}}\bigg|E^{\nu N}_{\nu}\right),
\end{equation}
where the two quantities satisfy $\log E^{\nu N}_{\gamma}
+ \log\!\left(m_{\pi}^2/ 4 E^{\nu N}_{\gamma}\right)
= 2 \log\!\left(m_{\pi}/2\right)$,
proving that $f$ is symmetric in $\log E^{\nu N}_{\gamma}$ around $m_{\pi}/2$.

In the laboratory frame, the Lorentz transformation in Eqs.~\eqref{eq:beta} and~\eqref{eq:transform} shifts the symmetry axis to
\begin{align}
    \label{eq:axis}
    E^{\rm lab}_{\rm sym}&=\sqrt{\frac{1+v}{1-v}}\frac{m_{\pi}}{2}=\sqrt{\frac{2E^{\rm lab}_{\nu}+m_N}{m_N}}\frac{m_{\pi}}{2} \\ \notag &\simeq \sqrt{\frac{E^{\rm lab}_{\nu}}{2m_{N}}}m_{\pi}.
\end{align}
which increases with $E^{\rm lab}_{\nu}$, as seen in Fig.~\ref{fig:results}.

Next, we analyze the upper and lower bounds of $E^{\rm lab}_{\gamma}$ in the photon probability density function $f(E^{\rm lab}_{\gamma}|E^{\rm lab}_{\nu})$. From Eq.~\eqref{eq:probability_formula_3}- Eq.~\eqref{eq:probability_formula_6}, we observe that, in addition to the lower bound $p^{\nu N}_{\rm low}$, the integral in Eq.~\eqref{eq:probability_formula_3} also has an upper bound, $p^{\nu N}_{\rm up}$, given by:
\begin{equation}
    \label{eq:up_bound}
    p^{\nu N}_{\rm up} = E^{\nu N}_{\nu}.
\end{equation}
Thus, if $p^{\nu N}_{\rm low}>p^{\nu N}_{\rm up}$, the integral in Eq.~\eqref{eq:probability_formula_3} vanishes and then $f=0$. The condition $p^{\nu N}_{\rm low}<p^{\nu N}_{\rm up}$ yields the allowed photon-energy window
\begin{equation}
    \label{eq:bound_gamma}
    \frac{\sqrt{(E_{\nu}^{\nu N})^2+m_{\pi}^2}-E^{\nu N}_{\nu}}{2}<E^{\nu N}_{\gamma}<\frac{\sqrt{(E_{\nu}^{\nu N})^2+m_{\pi}^2}+E^{\nu N}_{\nu}}{2}.
\end{equation}
In the limit $E_{\nu}^{\nu N}\gg m_{\pi}$, this reduces to
\begin{equation}
    \label{eq:bound_gamma_asymp}
    \frac{m_{\pi}^2}{4E^{\nu N}_{\nu}}<E^{\nu N}_{\gamma}<E^{\nu N}_{\nu}.
\end{equation}

Boosting to the laboratory frame, Eqs.~\eqref{eq:beta}–\eqref{eq:transform} give the upper bound
\begin{equation}
    \label{eq:gamma_up}
    E_{\rm up}^{\rm lab}=E^{\rm lab}_{\nu},
\end{equation}
i.e., the photon energy cannot exceed the incident (anti)neutrino energy. The lower bound becomes
\begin{align}
    \label{eq:gamma_low}
    &E_{\rm low}^{\rm lab} = \frac{m_{\pi}^2}{4E^{\rm lab}_{\nu}}\left(\frac{1+v}{1-v}\right) \\ \notag
    &=\frac{m_{\pi}^2}{4E^{\rm lab}_{\nu}}\frac{2E^{\rm lab}_{\nu}+m_N}{m_N}\simeq \frac{m_{\pi}^2}{2m_{N}},
\end{align}
which is independent of $E^{\rm lab}_{\nu}$ and numerically equals $9.7 \times 10^{-3}$ GeV, as clearly depicted in the left panels of Fig.~\ref{fig:results}.

Based on the energies of (anti)neutrinos recorded by IceCube and KM3NeT, which may originate from dense astrophysical environments, we investigate whether UHE incident (anti)neutrinos can produce UHE photons through scattering with nucleons. Our calculations show that, for an incident (anti)neutrino with energy $E^{\rm lab}_{\nu}$, the highest probability corresponds to producing photons with energies of approximately $0.10 \times\sqrt{E^{\rm lab}_\nu/\rm GeV}~ {\rm GeV}$. Meanwhile, for incident (anti)neutrinos with energies above 1 TeV, the probability of producing photons with energies exceeding 1 GeV is greater than 99\%.   For incident (anti)neutrinos with energies above 1 PeV, the probability of producing photons with energies exceeding 1 TeV is greater than 13\%. For incident 220 PeV (anti)neutrinos, the probability of producing photons with energies exceeding 1 TeV is greater than 94\%. 

In dense astrophysical environments such as the interior of Wolf–Rayet stars that serve as GRB progenitors, the secondary photons produced by this reaction are essentially unobservable if the interaction occurs in an optically thick region. However, if it takes place in an optically thin region, the resulting UHE photons may escape and become detectable. 
This motivates a concrete application to the preburst of GRB, which we quantify in the next section.

\section{Application to the Preburst Photons of GRB 221009A}
\label{sec:grb221009a}
\begin{figure*}[t]
    \centering
    \includegraphics[width=0.72\textwidth]{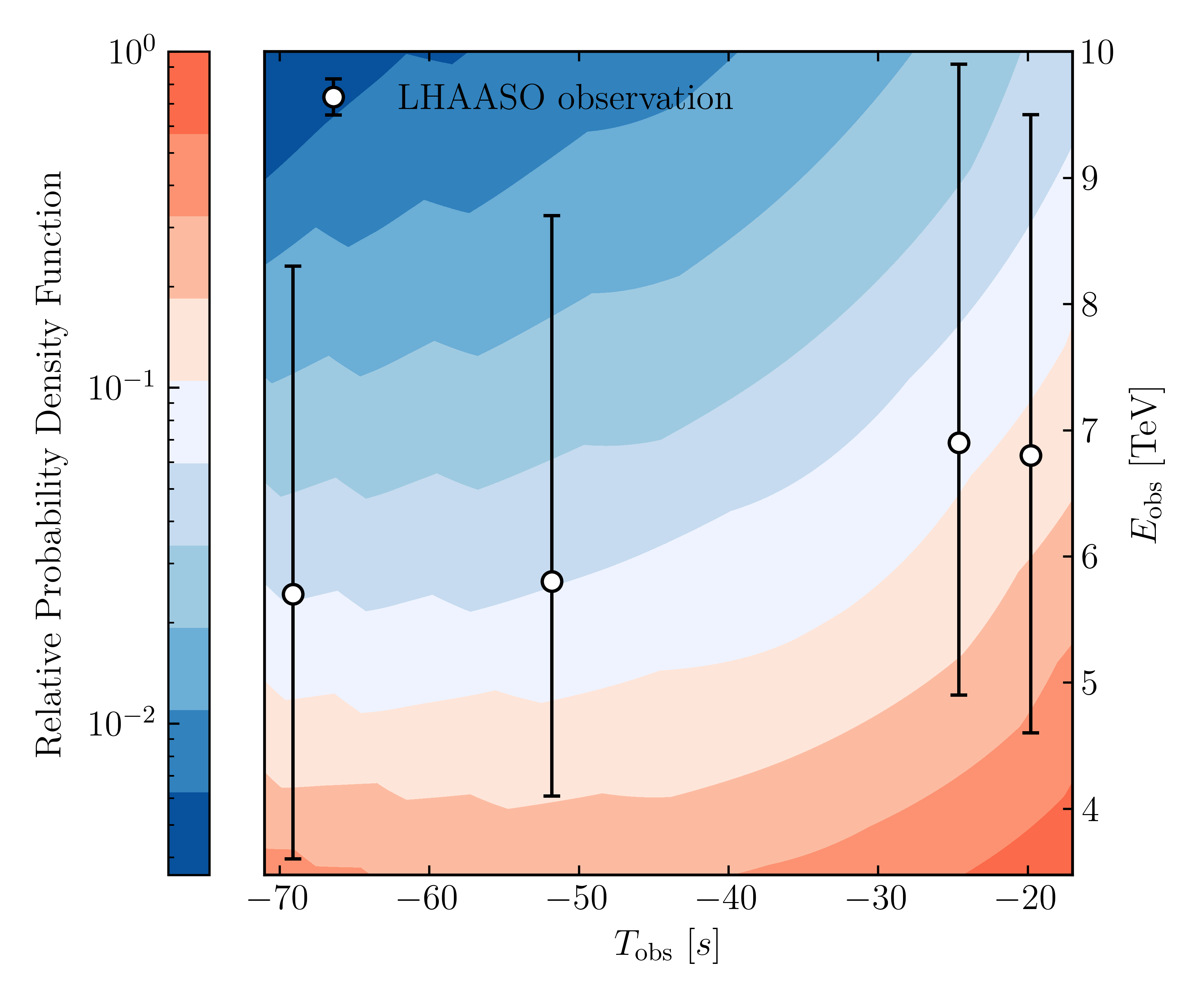}
    \caption{
        Relative two-dimensional probability density in the observed $(T_{\rm obs},E_{\rm obs})$ plane, normalized to its maximum within the displayed region. The benchmark parameters are $k=2.5$, $L_{0}=1$, $\rho_{0}=10$, and $R_{12}=20$. The data points with error bars indicate the reported LHAASO preburst photons of GRB~221009A. All points lie within the main high-probability region of the predicted distribution.
    }
    \label{fig:grb221009a_relative}
\end{figure*}

In this section, we apply this mechanism to a concrete astrophysical phenomenon that remains to be understood. A natural case study is the preburst emission of GRB~221009A. As shown in Fig.~\ref{fig:grb221009a_relative}, the KM2A detector of LHAASO observed four photons at TeV band prior to the main prompt emission, and even before the Fermi-GBM trigger time $t_0$, with their arrival times distributed between $t_0-69.1~{\rm s}$ and $t_0-19.8~{\rm s}$~\cite{LHAASA:2023pay,LHAASO:2023kyg,Liu:2024qbt}. These events have not yet received a widely accepted explanation within conventional prompt or afterglow frameworks~\cite{Stern:2023qyl}.

For the scenario considered here, neutrinos are produced while the jet is still propagating inside the progenitor. Photons generated in the same inner region are trapped due to the large optical depth, whereas neutrinos can escape efficiently. Such a precursor neutrino component is also supported by studies of GRB-associated neutrinos observed by IceCube~\cite{Huang:2019etr, Huang:2022xto}. These escaping neutrinos subsequently interact with protons and neutrons in the optically thin outer region of the hydrogen (H) envelope, producing $\pi^{0}$ and, through their decay, UHE photons via the mechanism described in Secs.~\ref{Calculation} and~\ref{Results}. 
As a consequence, these (anti)neutrino-produced photons can manifest as a separated precursor signal.

To quantify this picture, we adopt the jet-head propagation framework of Refs.~\cite{MeszarosWaxman2001,Razzaque2003}. In the H envelope, the Lorentz factor of the jet head is given by
\begin{equation}
    \Gamma_h = 3\frac{L_{52}^{1/4}}{r_{12}^{1/2}\rho_{-7}^{1/4}},
\end{equation}
where the dimensionless quantities are defined as
\begin{align}
    r_{12}&\equiv\frac{r}{10^{12}\ \mathrm{cm}},\\ \rho_{-7}&\equiv\frac{\rho}{10^{-7}\ \mathrm{g}\cdot\mathrm{cm}^{-3}},\\ L_{52}&\equiv \frac{L_{\rm iso}}{10^{52}\ \mathrm{erg}/\mathrm{s}}.
\end{align}
Here $r$ is the shock radius, $L_{\rm iso}$ is the isotropic jet luminosity, and $\rho$ denotes the density of the H envelope at radius $r$.

For the jet propagation, we assume a constant isotropic luminosity $L_{\rm iso}=10^{52}\,L_{0}\ \mathrm{erg\,s^{-1}}$ and a power-law density profile,
\begin{equation}
    \rho(r)=10^{-7}\rho_{0}\left(\frac{r}{10^{12}\ {\rm cm}}\right)^{-k}\ {\rm g\,cm^{-3}},
\end{equation}
where the index $k$ is typically in the range $2.0$--$2.5$~\cite{Bhattacharya:2022btx}.

Under these assumptions, the jet-head Lorentz factor and velocity become
\begin{align}
    \Gamma_{h}&=3\left(\frac{L_{0}}{\rho_{0}}\right)^{1/4} r_{12}^{(k-2)/4}, \label{eq:grb_gammah}\\
    v_{h}&=c\sqrt{1-\Gamma_{h}^{-2}}
    =c\sqrt{1-\frac{1}{A_{0}r_{12}^{(k-2)/2}}}, \label{eq:grb_vh}
\end{align}
where $A_{0}\equiv 9\left(L_{0}/\rho_{0}\right)^{1/2}$.

If a (anti)neutrino is produced when the jet head is at radius $r_{\rm gen}$, the corresponding lead time relative to photons emitted only after jet breakout is
\begin{align}
\label{eq:grb_delay}
    T&=\int_{r_{\rm gen}}^{R}\frac{\mathrm{d}r'}{v_h}-\int_{r_{\rm gen}}^{R}\frac{\mathrm{d}r'}{c}\\ \notag
    &=33~{\rm s}\int_{r_{\rm gen, 12}}^{R_{12}}
    \left[\frac{1}{\sqrt{1-A_{0}^{-1}(r_{12}^{\prime})^{-(k-2)/2}}}-1\right]\mathrm{d}r_{12}^{\prime},
\end{align}
where we have defined
\begin{equation}
    R_{12}\equiv \frac{R}{10^{12}\ {\rm cm}},\quad r_{\rm gen, 12} \equiv \frac{r_{\rm gen}}{10^{12}\ {\rm cm}}.
\end{equation}
Here $R$ is the outer radius of the H envelope.

Following Ref.~\cite{MeszarosWaxman2001}, the (anti)neutrino energy in the laboratory frame is
\begin{align}
    E_{\nu}^{\rm lab}&\simeq 2\left(\frac{1+z}{2}\right)^{-1}\Gamma_{h}^{3/2}r_{\rm gen,12}^{1/2}L_{0}^{-1/4}\ {\rm TeV} \notag\\
    &=18\,B_{0}\,r_{\rm gen,12}^{(3k-2)/8}\ {\rm TeV},
    \label{eq:grb_enu}
\end{align}
where $B_{0}\equiv L_{0}^{1/8}\rho_{0}^{-3/8}$, and in the second line we have used the redshift $z=0.151$ for GRB~221009A~\cite{Lesage:2023vvj}.

The generation probability of precursor (anti)neutrinos is derived to follow the standard $E_{\nu}^{-2}$ spectrum~\cite{MeszarosWaxman2001},
\begin{equation}
    \label{eq:grb_pgen}
    P_{\rm gen}(T)\propto \left[E_{\nu}^{\rm lab}(T)\right]^{-2}.
\end{equation}

\begin{figure}[t]
    \centering
    \includegraphics[width=0.48\textwidth]{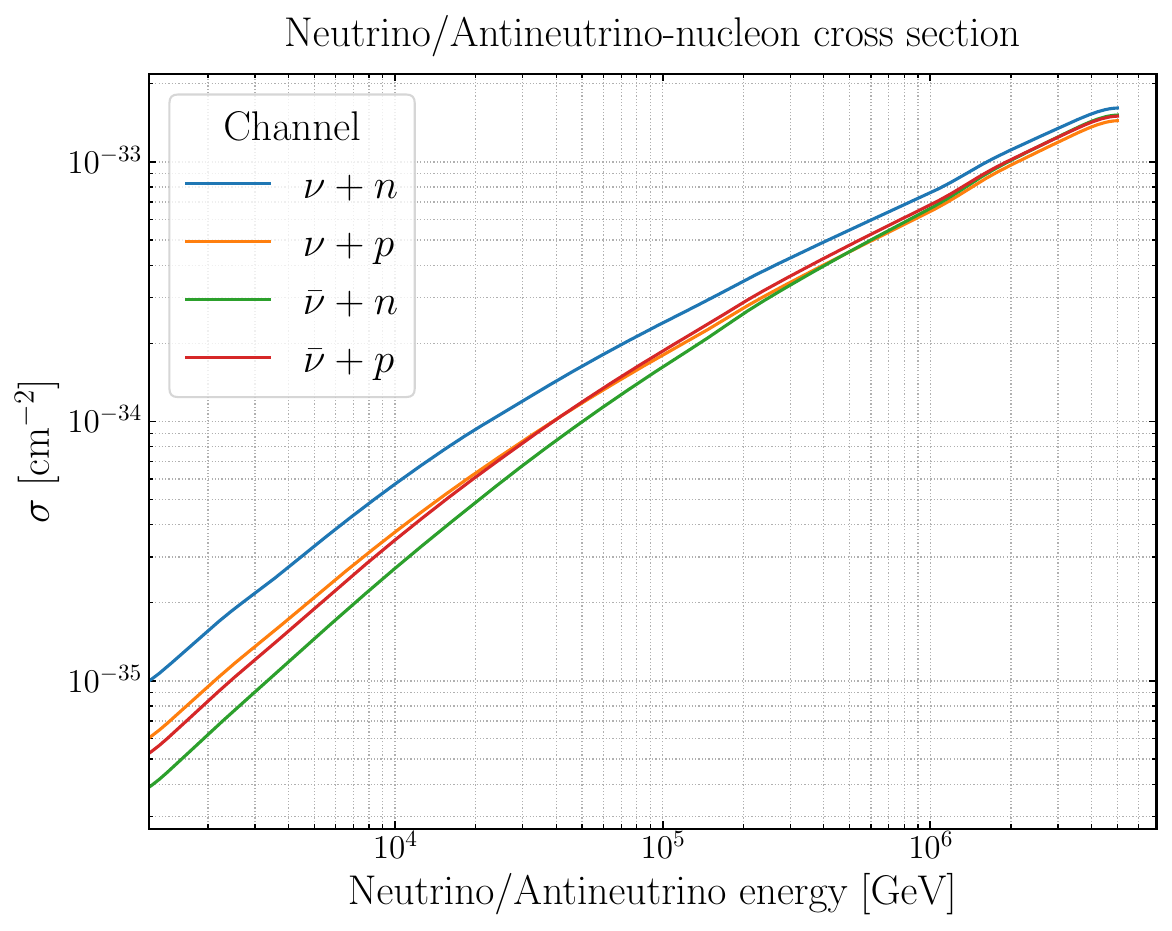}
    \caption{
        Total (anti)neutrino--nucleon cross sections as functions of the (anti)neutrino energy, including the channels $\nu+n$ (blue), $\nu+p$ (orange), $\bar{\nu}+n$ (green), and $\bar{\nu}+p$ (red). Here all final states are included, rather than only those leading to pion production.
    }
    \label{fig:cross_section}
\end{figure}

The probability that such a (anti)neutrino propagates to the optically thin region without further absorption is given by
\begin{align}
    \label{eq:grb_ppro}
    P_{\rm pro}^{(a)}(T)&\propto \frac{\exp\!\left[-\int_{r_{\rm gen}(T)}^{R}n(r^{\prime})\sigma_{a}\!\left(E_{\nu}^{\rm lab}(r^{\prime})\right)\mathrm{d}r^{\prime}\right]}{\left[R-r_{\rm gen}(T)\right]^{2}},\\ \notag
    &\propto \frac{e^{-6\rho_0\int_{r_{\rm gen,12}(T)}^{R_{12}}(r'_{12})^{-k} \sigma_{-28}(r'_{12})\mathrm{d}r'_{12}} }{[R_{12}-r_{\rm gen,12}(T)]^2}
\end{align}
where $a=\nu,\bar{\nu}$ labels neutrinos and antineutrinos, $n(r)$ is the nucleon number density,
\begin{equation}
    n(r)=6\times 10^{16}\ \mathrm{cm}^{-3}\ \rho_0\ \left(\frac{r}{10^{12}\ \mathrm{cm}}\right)^{-k},
\end{equation}
and $\sigma_{a}$ denotes the total $\nu N$ (or $\bar{\nu}N$) cross section obtained in this work (see Fig.~\ref{fig:cross_section}). The exponential factor in Eq.~\eqref{eq:grb_ppro} accounts for attenuation due to (anti)neutrino--nucleon interactions along the propagation path, while the denominator reflects the geometric dilution of the (anti)neutrino flux, which follows an inverse-square law with distance.

Finally, the observable joint distribution of precursor photons can be written as
\begin{align}
    \label{eq:grb_joint}
    &\mathcal{P}\!\left(T,E_{\gamma}^{\rm lab}\right)\\ \notag
    &\propto
    \sum_{a=\nu,\bar{\nu}}P_{\rm gen}(T)\,
    P_{\rm pro}^{(a)}(T)\,
    f_{a}\!\left(E_{\gamma}^{\rm lab}\big|E_{\nu}^{\rm lab}(T)\right),
\end{align}
where $f_{a}(E_{\gamma}^{\rm lab}|E_{\nu}^{\rm lab})$ is the photon energy distribution analyzed in Secs.~\ref{Calculation} and~\ref{Results}.
This expression provides a factorized description of the full process, combining the neutrino production, propagation, and conversion into observable photons. In the numerical application below, Eq.~\eqref{eq:grb_enu} implies that $E_{\nu}^{\rm lab}\sim \mathcal{O}(10)~\mathrm{TeV}$ throughout the relevant region of the envelope, consistent with the preburst observations of GRB~221009A.

In the numerical calculation, we adopt the benchmark parameters $L_{0}=1$, $\rho_{0}=10$, and $R_{12}=20$, and assume that the inner radius of the H envelope is $\sim 10^{11}\ \mathrm{cm}$. In this parameter regime, the total $\nu N$ and $\bar{\nu}N$ cross sections increase from $\sim 10^{-35}\ \mathrm{cm^{2}}$ to $\sim 10^{-34}\ \mathrm{cm^{2}}$, as shown in Fig.~\ref{fig:cross_section}.

\begin{figure}[t]
    \centering
    \includegraphics[width=\linewidth]{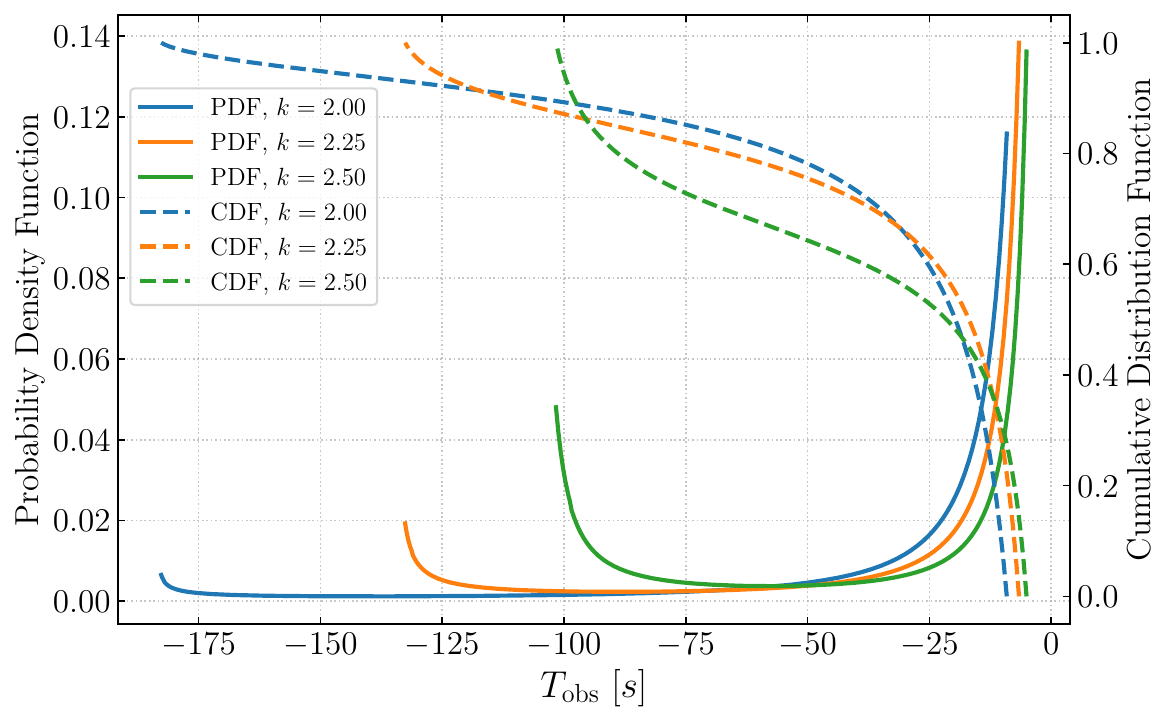}
    \caption{
        Lead-time distribution for GRB~221009A.
        Solid curves: normalized probability density of the lead time $T_{\rm obs}\equiv -T$ for $k=2.00$ (blue), $k=2.25$ (orange), and $k=2.50$ (green).
        Dashed curves: corresponding cumulative distribution functions.
        The benchmark parameters are $L_{0}=1$, $\rho_{0}=10$, and $R_{12}=20$.
    }
    \label{fig:grb_benchmark}
\end{figure}

The left and right panels of Fig.~\ref{fig:grb_benchmark} show the resulting lead-time distributions for different values of $k$ for PDF and CDF, respectively, where we define $T_{\rm obs}\equiv -T$ so that preburst photons correspond to negative observed times relative to the trigger time at $T_{\rm obs}=t_0=0~{\rm s}$. The maximum lead time is $T_{\max}\simeq 180~\mathrm{s}$. 
The distribution is strongly weighted toward short lead times. This behavior arises because a smaller $T$ corresponds to a larger production radius $r_{\rm gen}$ (see Eq.~\eqref{eq:grb_delay}), which in turn leads to a larger propagation probability $P_{\rm pro}^{(a)}(T)$ (see Eq.~\eqref{eq:grb_ppro}). 

At the same time, the distribution exhibits a mild enhancement at large $T$. This can be understood as follows: a larger $T$ corresponds to a smaller neutrino energy $E_{\nu}^{\rm lab}$ (see Eqs.~\eqref{eq:grb_delay} and~\eqref{eq:grb_enu}), and hence to a larger generation probability $P_{\rm gen}(T)$ (see Eq.~\eqref{eq:grb_pgen}). 

As a result of these competing effects, the distribution develops a long tail extending to $\mathcal{O}(10^{2})~\mathrm{s}$. This naturally allows for precursor signals appearing tens of seconds prior to the prompt emission in this scenario.

To facilitate a direct comparison with observational data, we define $E_{\rm obs}\equiv E^{\rm lab}_{\gamma}$ and focus on the LHAASO-relevant energy window $E_{\rm obs}=4$--$10~\mathrm{TeV}$. The resulting relative probability density, normalized to its maximum within the displayed window, is shown in Fig.~\ref{fig:grb221009a_relative}. The distribution exhibits a pronounced high-probability ridge extending across photon energies of a few TeV and lead times of several tens of seconds.

As shown in Fig.~\ref{fig:grb221009a_relative}, all four LHAASO preburst events fall within the predicted support of the distribution and overlap with the main high-probability region. This indicates that the (anti)neutrino-induced photon production mechanism studied here can quantitatively account for the observed preburst window of GRB~221009A. We emphasize that this mechanism is not necessarily unique. Rather, it provides a viable precursor channel in which (anti)neutrinos escape from the optically thick inner region and subsequently convert into photons in the outer optically thin envelope, leading to a precursor signal.

\section{Discussion  and Summary}
\label{Summary}
In this work, we investigated the production of high-energy photons via the neutrino–nucleon scattering process followed by $\pi^0$ decay. By calculating the neutrino–parton cross sections, incorporating nucleon PDFs and pion fragmentation functions, and evaluating the decay rates of $\pi^0$ into photons, we obtained the photon energy distributions for incident (anti)neutrinos with energies ranging from 1 TeV to 220 PeV. The results indicate that this process can efficiently produce UHE photons, with nearly identical outcomes for neutrinos and antineutrinos. 
We further analyzed the resulting photon energy spectra and found that for (anti)neutrinos with energies above 1 PeV, the probability of producing photons exceeding 1 TeV is greater than 13\%—a non-negligible fraction that supports potential observational relevance.

In dense astrophysical environments such as the jets of active galactic nuclei (AGN)~\cite{Dermer:2014vaa}, the internal shocks of GRBs~\cite{Murase:2013ffa}, and the outer layers of core-collapse supernovae~\cite{Bykov:2015nta, Chakraborty:2015sta, Petropoulou:2017ymv}, UHE neutrinos can be copiously produced through hadronic interactions of accelerated protons with ambient matter or radiation fields. If a fraction of these neutrinos traverse regions of moderate density, where the optical depth is low enough for photons to escape, the mechanism investigated in this work can operate efficiently. Under such conditions, UHE neutrinos may generate secondary photons via neutral pion decay following $\nu N$ scattering, adding a subdominant but potentially observable component to the overall gamma-ray flux.

Our results quantitatively demonstrate that this process, while contributing only a small fraction of the total emission, could nonetheless produce detectable high-energy photons under favorable conditions. 
As a concrete example, the observed preburst of GRB~221009A, characterized by photon energies of a few TeV and lead times of several tens of seconds~\cite{Liu:2024qbt,Song:2024and,Song:2025qej,Song:2025myx, Song:2025ksi}, can be naturally accommodated within our framework. By combining the neutrino production, propagation, and conversion processes, our model provides a self-consistent quantitative description of both the energy scale and the temporal structure of the preburst signal. 
In this sense, the mechanism developed in this work is not merely a theoretical possibility, but can be applied to a concrete astrophysical scenario.

\reply{We have also estimated the luminosity of neutrinos $L_\nu$, and that of the parent protons $L_p$, required to reproduce the LHAASO preburst events. For the benchmark neutrino--nucleon cross section and envelope parameters considered in this work, the required precursor-neutrino luminosity lies within a reasonable range for an exceptionally energetic burst such as GRB~221009A. The corresponding parent-proton luminosity depends mainly on the neutrino-production efficiency, defined by $E_{\nu,{\rm iso}}^{\rm req}=\xi_\nu E_{p,{\rm iso}}^{\rm req}$. For typical values of $\xi_\nu$ adopted in GRB-neutrino estimates, the required isotropic-equivalent parent-proton energy remains comparable to or below the available isotropic-equivalent energy budget of GRB~221009A. This normalization check therefore supports the energetic consistency of the proposed preburst neutrino-induced photon scenario within the benchmark parameter region considered here.
}

\reply{The associated multi-wavelength electromagnetic emission is also expected to remain consistent with observations. The photons produced together with the precursor neutrinos are generated before jet breakout, inside the dense stellar envelope, where both the primary TeV photons and the lower-energy cascade photons are efficiently trapped. The absorbed electromagnetic energy is therefore released only on a much longer diffusion timescale and is softened by repeated scattering and partial thermalization, rather than escaping promptly as an X-ray or GeV counterpart to the LHAASO preburst photons. 
In addition, Swift/XRT did not observe the GRB field during the preburst time window of GRB~221009A~\cite{Williams:2023sfk}. Although Fermi-GBM and Fermi-LAT observed GRB~221009A after the trigger during the prompt phase, the expected escaping cascade flux associated with the preburst phase is far below the relevant triggering or detection thresholds of these instruments~\cite{Meegan:2009qu,Nava:2016onl}. Thus, while a full cascade and radiation-transfer calculation is beyond the scope of this work, our order-of-magnitude estimates indicate that the benchmark scenario is compatible with the available multi-wavelength observations.}

This work establishes a new quantitative pathway linking high-energy neutrino interactions to gamma-ray emission, expanding our understanding of multi-messenger astrophysics in extreme environments. The consistency between neutrino and antineutrino photon production further simplifies observational predictions, highlighting the potential of this process to constrain neutrino origins and astrophysical environment properties through future multi-wavelength and multi-messenger surveys.

\emph{\textbf{Acknowledgements}.---}
We thank the anonymous referee for valuable comments and suggestions. This work is supported by the National Natural Science Foundation of China under grants No. 12335006 and No. 12675108.  This work is also supported by the High-performance Computing Platform of Peking University.

\bibliography{scibib}

\appendix
\section{Amplitude Square of the Neutrino-Parton Scattering}
\label{appendix:amplitude_square}
In this appendix we collect the squared amplitudes for all neutrino–parton and antineutrino–parton scattering processes. Owing to the generation structure of the electroweak interaction, it is sufficient to present the results for the first generation. The corresponding expressions for second- and third-generation leptons and quarks are obtained by replacing the masses and the relevant CKM matrix elements in the formulas below.

\begin{figure*}[t]
	\centering
	\includegraphics[width=0.45\linewidth]{figures/Process01.pdf}
	\includegraphics[width=0.45\linewidth]{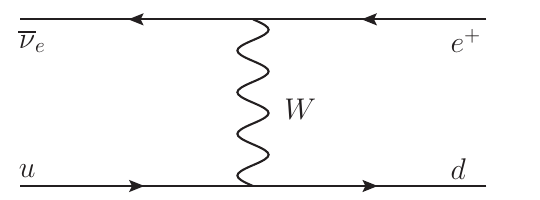}
	\includegraphics[width=0.45\linewidth]{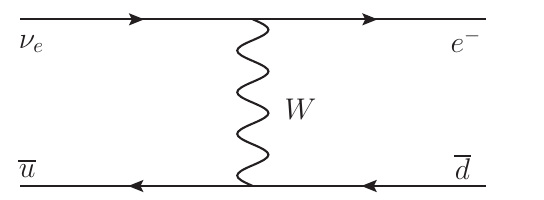}
	\includegraphics[width=0.45\linewidth]{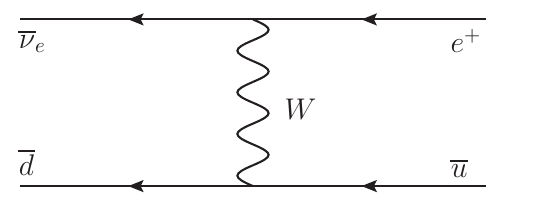}
	\includegraphics[width=0.45\linewidth]{figures/Process05.pdf}
	\includegraphics[width=0.45\linewidth]{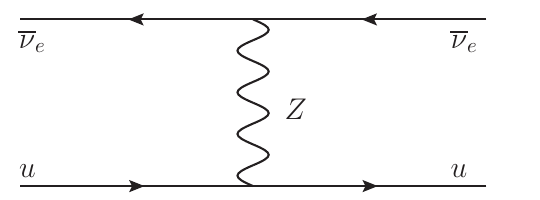}
	\includegraphics[width=0.45\linewidth]{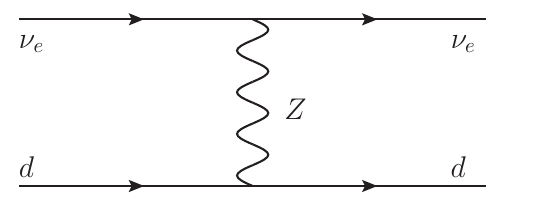}
	\includegraphics[width=0.45\linewidth]{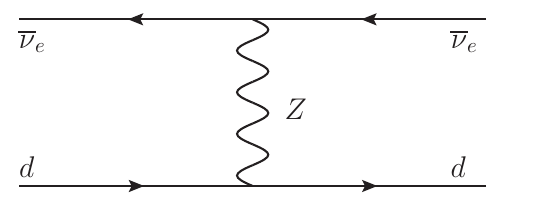}
	\includegraphics[width=0.45\linewidth]{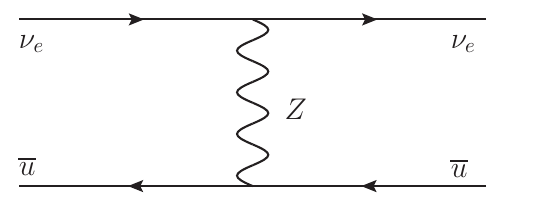}
	\includegraphics[width=0.45\linewidth]{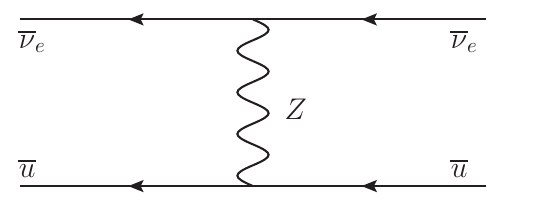}
	\includegraphics[width=0.45\linewidth]{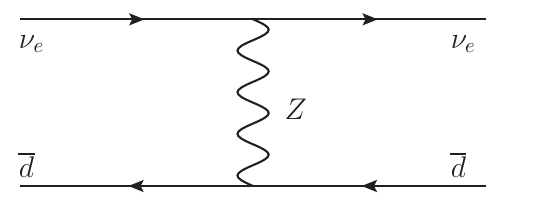}
	\includegraphics[width=0.45\linewidth]{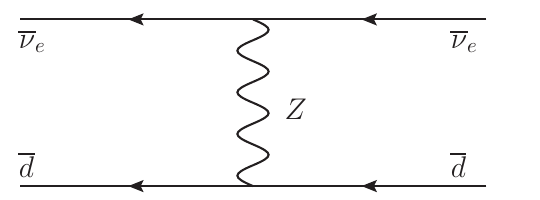}
	\caption{
		Feynman diagrams for neutrino–parton and antineutrino–parton scattering.  
		Diagrams in the left column correspond to neutrino-induced processes, while those in the right column correspond to antineutrino-induced processes.
	}
	\label{fig:Feynman_full}
\end{figure*}

The tree-level Feynman diagrams for all processes are shown in Fig.~\ref{fig:Feynman_full}, where the left column corresponds to neutrino-induced channels and the right column to antineutrino-induced channels.

The squared amplitudes for the processes with an initial neutrino are
\begin{align}
	\label{eq:process_one_new}
	&|\mathcal{M}_{d,u}|^2=|V_{ud}|^2\frac{e^4}{\sin^4 \theta_w} \\ \notag
	&\times \frac{P_{\nu} \cdot P_{i}\left(m_d^2-m_e^2-m_u^2+2P_{\nu} \cdot P_{i}\right)}{\left(m_{W}^2-m_e^2+2P_{\nu} \cdot P_{l} \right)^2},
\end{align}
\begin{align}
	\label{eq:process_three_new}
	&|\mathcal{M}_{\overline{u},\overline{d}}|^2=|V_{ud}|^2\frac{e^4}{\sin^4 \theta_w} \left(P_{\nu} \cdot P_{i}-P_{\nu} \cdot P_{l}\right) \\ \notag
	&\times \frac{\left(m_e^2+m_u^2-m_d^2+2P_{\nu} \cdot P_{i}-2P_{\nu} \cdot P_{l}\right)}{\left(m_{W}^2-m_e^2+2P_{\nu} \cdot P_{l} \right)^2},
\end{align}
\begin{align}
	\label{eq:process_five_new}
	&|\mathcal{M}_{u,u}|^2=\frac{e^4}{\cos^4 \theta_w} \frac{1}{18\left(m_{Z}^2+2P_{\nu} \cdot P_{l}\right)^2} \\ \notag
	&\times \bigg[4 P_{\nu} \cdot P_{l} \left(m_{u}^2 \left(3 \csc^2 \theta_w -4\right)-8 P_{\nu} \cdot P_{i} \right)  \\ \notag
	&+ \left(9 \csc^4 \theta_w -24 \csc^2 \theta_w + 32\right) \left(P_{\nu} \cdot P_{i}\right)^2 + 16 \left(P_{\nu} \cdot P_{l}\right)^2\bigg],
\end{align}
\begin{align}
	\label{eq:process_six_new}
	&|\mathcal{M}_{d,d}|^2=\frac{e^4}{\cos^4 \theta_w}\frac{1}{18\left(m_{Z}^2+2P_{\nu} \cdot P_{l}\right)^2} \\ \notag
	&\times \bigg[2 P_{\nu} \cdot P_{l} \left(m_d^2\left(3 \csc^2 \theta_w -2\right)-4 P_{\nu} \cdot P_{i} \right)  \\ \notag
	&+ \left(P_{\nu} \cdot P_{i}\right)^2 \csc^4 \theta_w \left(2 \cos(2 \theta_w)+\cos(4 \theta_w)+6\right) \\ \notag
	&+ 4 \left(P_{\nu} \cdot P_{l}\right)^2 \bigg],
\end{align}
\begin{align}
	\label{eq:process_seven_new}
	&|\mathcal{M}_{\overline{u},\overline{u}}|^2=\frac{e^4}{\cos^4 \theta_w \sin^4 \theta_w}\frac{1}{18\left(m_{Z}^2+2P_{\nu} \cdot P_{l}\right)^2}\\ \notag
	& \times \bigg[ P_{\nu} \cdot P_{l} (4 \cos^2 \theta_w -1 ) ( 4m_u^2 \sin^2 \theta_w \\ \notag
	&+ \left(4 \cos^2 \theta_w -1\right) P_{\nu} \cdot P_{l} ) \\ \notag 
	& + \left(P_{\nu} \cdot P_{i} \right)^2 \left(4 \cos(4 \theta_w)-4 \cos (2\theta_w)+9\right) \\ \notag 
	&-2 \left(P_{\nu} \cdot P_{i} \right) \left( P_{\nu} \cdot P_{l} \right) \left( 4 \cos^2 \theta_w -1 \right)^2 \bigg],
\end{align}

\begin{align}
	\label{eq:process_eight_new}
	& |\mathcal{M}_{\overline{d},\overline{d}}|^2=\frac{e^4}{\cos^4 \theta_w \sin^4 \theta_w}\frac{1}{18\left(m_{Z}^2+2P_{\nu} \cdot P_{l}\right)^2}\\ \notag
	&\times \bigg[ P_{\nu} \cdot P_{l} \left(2 \cos^2 \theta_w + 1\right) (2 m_d^2 \sin^2 \theta_w  \\ \notag
	&+ P_{\nu} \cdot P_{l} \left(2\cos^2 \theta_w +1\right)) \\ \notag
	& +  \left( P_{\nu} \cdot P_{i} \right)^2 \left(\cos (4 \theta_w)+2 \cos(2 \theta_w)+6\right)  \\ \notag
	& -2 \left(P_{\nu} \cdot P_{i} \right) \left( P_{\nu} \cdot P_{l} \right) \left(2 \cos^2 \theta_w +1\right)^2 \bigg].
\end{align}
The squared amplitudes for the processes with an initial antineutrino are
\begin{align}
	\label{eq:process_two_new}
	&|\mathcal{M}_{u,d}|^2 = |V_{ud}|^2\frac{e^4}{\sin^4 \theta_w} \left(P_{\nu} \cdot P_{i}-P_{\nu} \cdot P_{l}\right) \\ \notag
	&\times \frac{\left(m_e^2+m_u^2-m_d^2+2P_{\nu} \cdot P_{i}-2P_{\nu} \cdot P_{l}\right)}{\left(m_{W}^2-m_e^2+2P_{\nu} \cdot P_{l} \right)^2},
\end{align}
\begin{align}
	\label{eq:process_four_new}
	&|\mathcal{M}_{\overline{d},\overline{u}}|^2 = |V_{ud}|^2\frac{e^4}{\sin^4 \theta_w} \\ \notag
	&\times \frac{P_{\nu} \cdot P_{i}\left(m_d^2-m_e^2-m_u^2+2P_{\nu} \cdot P_{i}\right)}{\left(m_{W}^2-m_e^2+2P_{\nu} \cdot P_{l} \right)^2},
\end{align}
\begin{align}
	\label{eq:process_nine_new}
	&|\mathcal{M}_{u,u}|^2 =\frac{e^4}{\cos^4 \theta_w \sin^4 \theta_w}\frac{1}{18\left(m_{Z}^2+2P_{\nu} \cdot P_{l}\right)^2}\\ \notag
	& \times \bigg[ P_{\nu} \cdot P_{l} (4 \cos^2 \theta_w -1 ) ( 4m_u^2 \sin^2 \theta_w \\ \notag
	&+ \left(4 \cos^2 \theta_w -1\right) P_{\nu} \cdot P_{l} ) \\ \notag 
	& + \left(P_{\nu} \cdot P_{i} \right)^2 \left(4 \cos(4 \theta_w)-4 \cos (2\theta_w)+9\right) \\ \notag 
	&-2 \left(P_{\nu} \cdot P_{i} \right) \left( P_{\nu} \cdot P_{l} \right) \left( 4 \cos^2 \theta_w -1 \right)^2 \bigg],
\end{align}
\begin{align}
	\label{eq:process_ten_new}
	&|\mathcal{M}_{d,d}|^2=\frac{e^4}{\cos^4 \theta_w \sin^4 \theta_w}\frac{1}{18\left(m_{Z}^2+2P_{\nu} \cdot P_{l}\right)^2}\\ \notag
	&\times \bigg[ P_{\nu} \cdot P_{l} \left(2 \cos^2 \theta_w + 1\right) (2 m_d^2 \sin^2 \theta_w  \\ \notag
	&+ P_{\nu} \cdot P_{l} \left(2\cos^2 \theta_w +1\right)) \\ \notag
	& +  \left( P_{\nu} \cdot P_{i} \right)^2 \left(\cos (4 \theta_w)+2 \cos(2 \theta_w)+6\right)  \\ \notag
	& -2 \left(P_{\nu} \cdot P_{i} \right) \left( P_{\nu} \cdot P_{l} \right) \left(2 \cos^2 \theta_w +1\right)^2 \bigg],
\end{align}
\begin{align}
	\label{eq:process_eleven}
	&|\mathcal{M}_{\overline{u},\overline{u}}|^2=\frac{e^4}{\cos^4 \theta_w} \frac{1}{18\left(m_{Z}^2+2P_{\nu} \cdot P_{l}\right)^2} \\ \notag
	&\times \bigg[4 P_{\nu} \cdot P_{l} \left(m_{u}^2 \left(3 \csc^2 \theta_w -4\right)-8 P_{\nu} \cdot P_{i} \right)  \\ \notag
	&+ \left(9 \csc^4 \theta_w -24 \csc^2 \theta_w + 32\right) \left(P_{\nu} \cdot P_{i}\right)^2 + 16 \left(P_{\nu} \cdot P_{l}\right)^2\bigg],
\end{align}
\begin{align}
	\label{eq:process_twelve}
	&|\mathcal{M}_{\overline{d},\overline{d}}|^2= \frac{e^4}{\cos^4 \theta_w}\frac{1}{18\left(m_{Z}^2+2P_{\nu} \cdot P_{l}\right)^2} \\ \notag
	&\times \bigg[2 P_{\nu} \cdot P_{l} \left(m_d^2\left(3 \csc^2 \theta_w -2\right)-4 P_{\nu} \cdot P_{i} \right)  \\ \notag
	&+ \left(P_{\nu} \cdot P_{i}\right)^2 \csc^4 \theta_w \left(2 \cos(2 \theta_w)+\cos(4 \theta_w)+6\right) \\ \notag
	&+ 4 \left(P_{\nu} \cdot P_{l}\right)^2 \bigg],
\end{align}
Finally, substituting Eqs.~\eqref{inner_product_formula_one}, \eqref{inner_product_formula_two}, and \eqref{eq:cos_thetha_two} into the expressions above, and neglecting lepton and parton masses, we are able to obtain the final forms of $|\mathcal{M}_{i,i'}|^{2}$ in the $\nu N$-frame.

\end{document}